\documentclass[10pt,a4paper]{article}
\usepackage{makeidx}
\usepackage{amssymb}
\usepackage{amsmath}
\usepackage{graphicx}
\usepackage{graphics}
\usepackage{times}
\usepackage[T1]{fontenc}
\usepackage[thinspace,thinqspace,squaren]{SIunits}
\usepackage{graphicx}
\usepackage{amsthm}
\usepackage{color}
\usepackage{hyperref}
\usepackage{multirow}

\usepackage{ctable}
\allowdisplaybreaks

\usepackage{algorithm}
\usepackage{algorithmic}

\onecolumn

\begin{document}
	\date{}
	
	\title{Autonomous Take-Off and Flight of a Tethered Aircraft for Airborne Wind Energy}
	
	
	\author{Lorenzo~Fagiano,
		Eric Nguyen-Van,
		Felix Rager,
		Stephan Schnez,
		and Christian Ohler%
		\thanks{This is the pre-print of a paper accepted for publication on the IEEE Transactions on Control systems Technology  and is subject to IEEE Copyright.  doi: 10.1109/TCST.2017.2661825. L. Fagiano is with Politecnico di Milano, Dip. di Elettronica, Informazione e Bioingegneria, 20133 Milano, Italy. E. Nguyen-Van, F. Rager, S. Schnez and C. Ohler are with ABB Switzerland Ltd., Corporate Research, 5405 Baden-D\"{a}ttwil - Switzerland. E-mail addresses: lorenzo.fagiano@polimi.it, \{eric-nguyen-van $|$ felix.rager $|$ stephan.schnez $|$ christian.ohler\}@ch.abb.com.}
		\thanks{Corresponding author: Lorenzo Fagiano.}}
	\maketitle
	\begin{abstract}
		A control design approach to achieve fully autonomous take-off and flight maneuvers with a tethered aircraft is presented and demonstrated in real-world flight tests with a small-scale prototype. A ground station equipped with a controlled winch and a linear motion system accelerates the aircraft to take-off speed and controls the tether reeling in order to limit the pulling force. This setup corresponds to airborne wind energy systems with ground-based energy generation and rigid aircrafts. A simple model of the aircraft's dynamics is introduced and its parameters are identified from experimental data. A model-based, hierarchical feedback controller is then designed, whose aim is to manipulate the elevator, aileron and propeller inputs in order to stabilize the aircraft during the take-off and to achieve figure-of-eight flight patterns parallel to the ground. The controller operates in a fully decoupled mode with respect to the ground station. Parameter tuning and stability/robustness aspect are discussed, too. The experimental results indicate that the controller is able to achieve satisfactory performance and robustness, notwithstanding its simplicity, and confirm that the considered take-off approach is technically viable and solves the issue of launching this kind of airborne wind energy systems in a compact space and at low additional cost.
	\end{abstract}

	\section{Introduction}\label{S:introduction}


	Airborne Wind Energy (AWE) systems aim at extracting energy from the wind by using autonomous tethered aircrafts, kept aloft by either aerodynamic or aerostatic lift \cite{Ahrens2013,Fagiano2012}. By using light and flexible tethers instead of a massive structure, like the tower of wind turbines, AWE technologies feature relatively low capital costs and can operate above 200$\,$m from the ground, where the wind is generally faster and more consistent.  
	The price paid for these advantages is a higher system complexity, which makes the development of AWE difficult. At the current stage, power production has been studied theoretically and assessed experimentally by several companies and academic research groups worldwide, for different AWE concepts \cite{Loyd80,Ilzhoefer2007,Fagiano2010b,Canale2010,VanderLind2013,Vlugt2013,Vermillion2014,Erhard2015,Zgraggen2016}.
	
	However, system reliability and robustness during power generation has not been demonstrated yet, not even on a small scale, at least judging from the publicly available literature.  Due to the complex  dynamics of AWE systems, involving the interaction of ground-based equipments with long, flexible tether(s) and with the aircraft immersed in an uncertain and variable wind field, successful simulations and small experiments do not represent a significant proof of long-term reliability and performance: only an experimental validation carried over several months can provide enough confidence in any AWE system to really foster its industrial development, and these long-term tests have not been carried out or documented yet. In turn, a prerequisite for such a long-term experimental validation is the development of a viable (technically and economically) approach for autonomous take-off and landing in compact space. For AWE systems that employ the so-called ``pumping'' or ``yoyo'' operation, i.e. where the tether is repeatedly reeled-out under high load from a winch installed on the ground and reeled-in under low load, this important feature appears to be still missing, again judging from the scientific literature. In fact, while for the power production phase there is a consensus on how the system shall operate, the related modeling and control aspects are relatively well-understood and experimental results have been reported \cite{Ruiterkamp2013,Fagiano2014,Erhard2015,Fechner2015,Zgraggen2016}, for take-off and landing it is not even clear what approach to pursue. Consequently, there appears to be a lack of understanding of what are the most relevant problems involved, and there are no documented  experimental results in the literature.
	
	There is therefore a need for research and development efforts targeted at studying and demonstrating take-off and landing strategies for pumping AWE systems, from first-principle, conceptual analyses down to the specific engineering problems that need to be solved during implementation.
	
	In this paper, we present a contribution in this direction, focusing our attention on the take-off phase of systems that employ rigid aircrafts. 
	In this context, outside the scientific literature there is evidence of autonomous take-off operation via a winch launch \cite{ampyx},
	which however requires
	a significant space in all directions in
	order to adapt to the wind conditions. Another take-off approach that has been practically demonstrated employs vertical-axis propellers to slowly lift the aircraft to a safe altitude before shifting to the power production phase \cite{twingtec}. In this case, the required land occupation is rather small, however the additional onboard mass might compromise the system's performance during power generation. Within the scientific literature, the take-off of pumping AWE generators has been addressed only to a limited extent, mainly considering a rotational take-off approach \cite{ZaGD13,Bont10}. In recent contributions \cite{Fagiano2016b,Fagianoa} we compared different take-off approaches on the basis of qualitative and quantitative criteria, and concluded that a linear approach is among the most promising solutions, with a good tradeoff between ground-based equipments, land occupation and additional onboard mass. In such an approach, the aircraft is accelerated to take-off speed over a short distance by a motor installed on the ground and then it climbs to a safe altitude using relatively small onboard propellers. At a recent AWE conference, the company Ampyx Power also announced the development of a similar approach for their system \cite{Kruijff2015}.
	
	The main contribution of this paper is to prove experimentally such a linear take-off concept on a small scale, with fully autonomous operation. Additional contributions are represented by the modeling and control design steps that we carried out in order to achieve this result. 
	The coordination between the ground station and the aircraft is essential to achieve a successful take-off. Here, we adopt a completely decoupled strategy, where the controller of the ground station and the one on the glider do not exchange any information. Rather, the ground station controller exploits a local, indirect measurement of the tether force to adjust the reeling speed. 
	For the sake of brevity, this control algorithm is only briefly described here; the full details are presented in 
	\cite{Fagiano2016submitted}. Instead, the focus of the present paper is on the design of the onboard control system and on the results of the autonomous take-off tests. We first introduce a simple model of the aircraft and identify its parameters using experimental data. Then, we use the model to design a hierarchical controller 
	able to achieve autonomous take-off, transition to figure-of-eight patterns and flight along such patterns at roughly constant elevation. We finally present and discuss  experimental results. Albeit the tests have been performed on a small-scale prototype, the control approach and most of the findings can be easily carried over to larger system sizes. 
	We conclude with some final remarks and indicate the potential  improvements that can be subject of future research. 
	It is important to point out that our whole research activity towards autonomous take-off and flight has been conceived as a proof-of-concept study rather than a final, optimized solution. To the best of our knowledge, this is nevertheless the first contribution in the scientific literature that provides the full details and the experimental validation of a control approach for the take-off of a tethered aircraft in compact space, to be used for AWE systems.
	
	The paper is organized as follows. Section \ref{S:proto_layout} presents the system layout, the features of the employed experimental setup and the considered control objectives. Section \ref{S:design} describes the modeling, identification and control design aspects. The experimental results are presented in section \ref{S:results}, and conclusions are drawn in section \ref{S:conclusions}.
	
	\section{System description and control objectives}\label{S:proto_layout}
	
\begin{figure}[!tbh]
		\begin{center}
			\includegraphics[width=12cm]{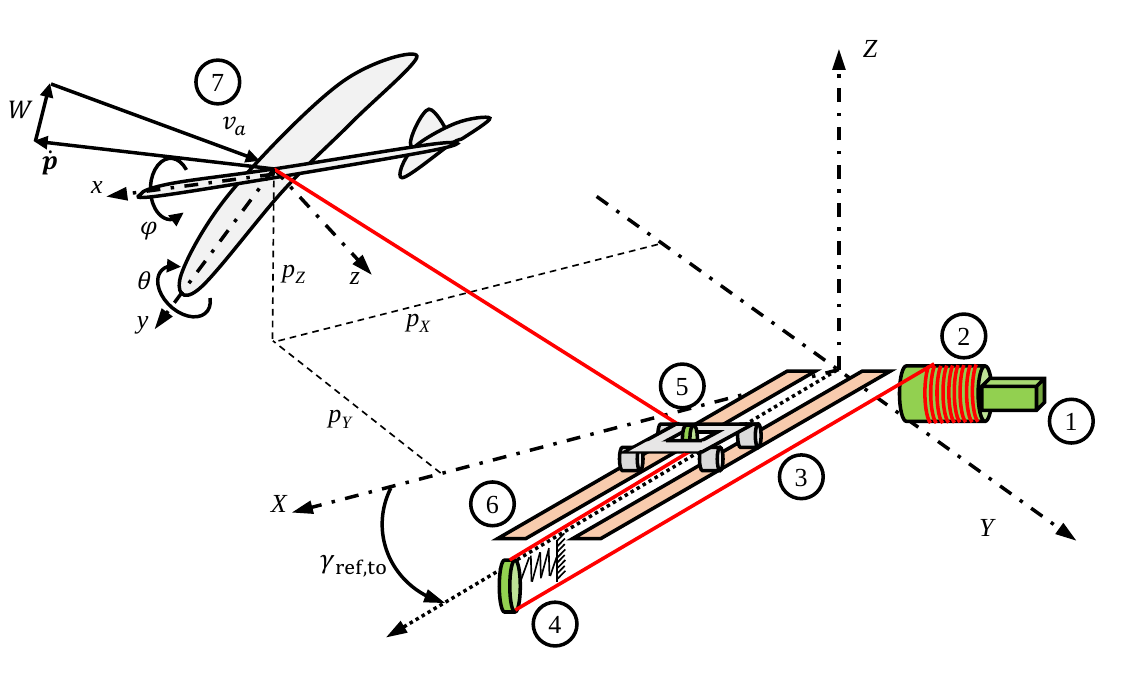}
			\caption{Sketch of the considered system, showing: 1. the winch motor, 2. the winch, 3. the tether connecting the winch to the aircraft, 4. the tether tensioning system, 5. the slide, 6. the rails, 7. the aircraft. The inertial coordinate system $(X,Y,Z)$ and the local one $(x,y,z)$ are also shown, together with the aircraft position $\boldsymbol{p}$ and velocity $\dot{\boldsymbol{p}}$, the roll and pitch rotations $\varphi,\,\theta$,  the  wind velocity vector relative to ground, $\boldsymbol{W}$, the resulting apparent wind velocity $\boldsymbol{v_a}$, and the angle $\gamma_{\text{ref,to}}$ between the take-off direction (dotted arrow) and the $X$-axis.}
			\label{F:sys_pic}
		\end{center}
	\end{figure}

\begin{figure}[!tbh]
		\begin{center}
			\includegraphics[width=12cm]{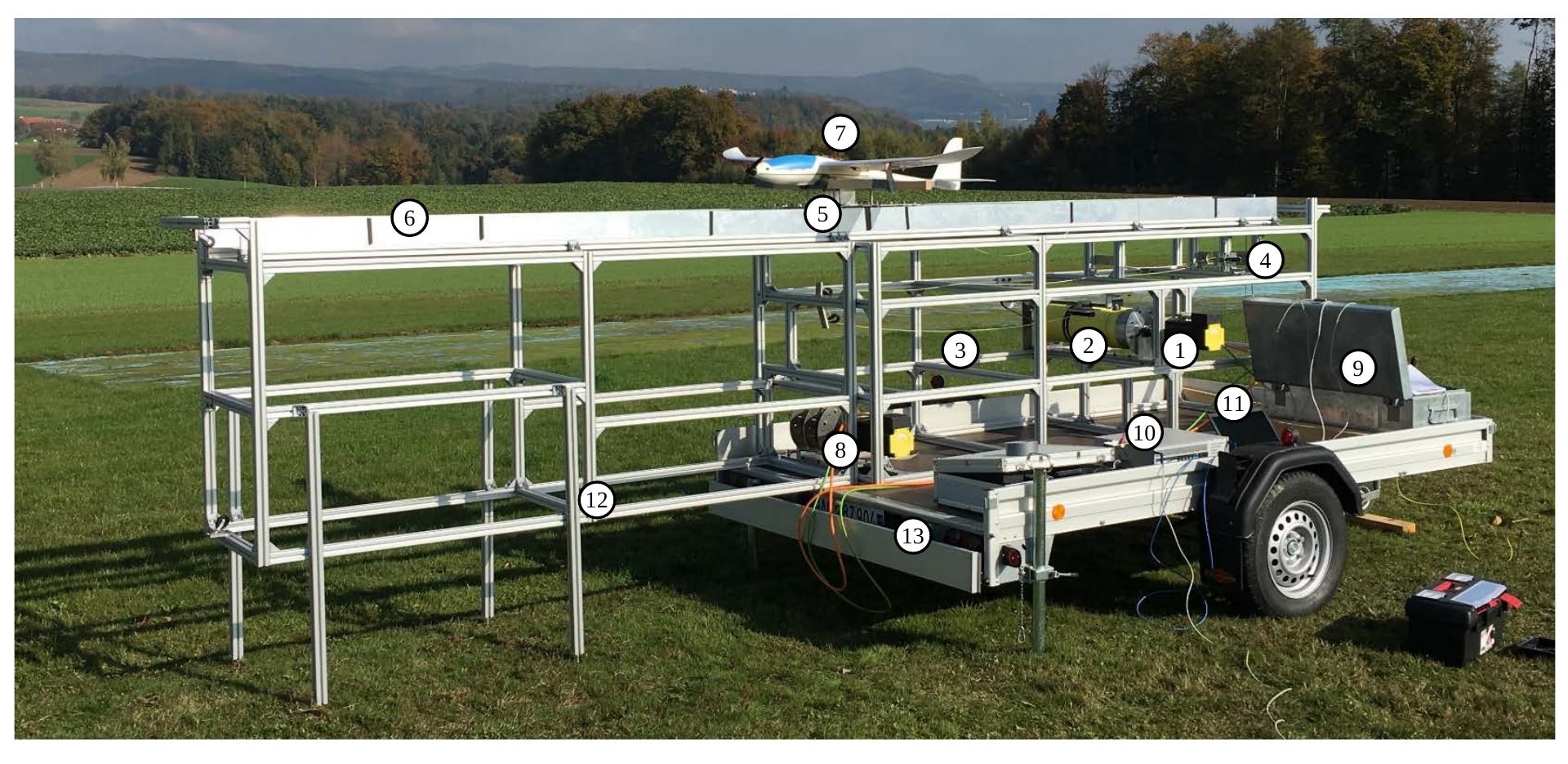}
			\caption{Picture of the small-scale prototype built at ABB Corporate Research. The numbers 1. to 7. correspond to the components also depicted in Fig. \ref{F:sys_pic}. In addition, the picture shows: 8. the slide motor and drum, 9. the box containing power supply, inverter, and drives, 10. the real-time machine for data-logging and high-level control of the ground station, 11. the laptop used to interface with the drives and the real-time machine.}
			\label{F:proto_pic}
		\end{center}
	\end{figure}

	The considered system is composed of a ground station connected to a rigid aircraft by a single tether, see Fig. \ref{F:sys_pic} for a sketch and Fig. \ref{F:proto_pic} for a picture of the employed small-scale prototype. We first describe the two main subsystems, i.e. the ground station and the aircraft, followed by a brief overview of the ground station control strategy. We finally present the desired system operation and the consequent control objectives for the aircraft's controller.
	
	\subsection{Ground station}\label{SS:GS}
	
	Referring to Figs. \ref{F:sys_pic}-\ref{F:proto_pic}, the ground station is equipped with a first motor/generator (``winch motor'') connected to a winch that stores the tether attached to the aircraft. In a power-generating system, the winch motor also acts as generator during the power production phase. 
	Before leaving the ground station, the tether passes through a series of pulleys, one of which is attached to a mass-spring mechanism used to reduce the stiffness of the link between the aircraft and the ground station. The last pulley 
	is installed on a slide able to move on linear rails. The slide is used to accelerate the aircraft up to take-off speed. The slide position is controlled by a second motor (``slide motor'') installed on the ground station. In this way, most of the power injected during the take-off maneuver is provided by the slide motor and the resulting required space can be quite compact: in our prototype, only about 3.5 m of travel distance are needed to accelerate to about 9 m/s and then brake the slide. 
	Finally, the ground station also comprises a power supply, the drives to control the two motors, a real-time machine for control and data-acquisition, a human-machine interface and the mechanical frame that supports all the mentioned components. In our prototype, the power supply is composed of two 12V batteries connected in parallel and  a DC-AC inverter providing a 220 VAC, 3 kW single-phase line to the drives and to all the measurement and control hardware. The main features of our small-scale prototype are reported in Table \ref{T:gs_data}, while the full details are given in \cite{Fagiano2016submitted}.
	
	\begin{table}[htb!]
		\caption{Main technical features of the employed prototype.}
		\label{T:gs_data}
		\centering
		\begin{tabular}{|l|r|r|}\hline
			\multicolumn{3}{|l|}{\textbf{Ground station}}\\\hline\hline
			Winch and slide motors' maximum continuous torque & 13 & Nm\\\hline
			Winch and slide motors' peak torque & 26 & Nm\\\hline
			Winch and slide motors' rated speed & 208 & rad/s\\\hline
			Tether length & 150 & m\\\hline
			Rails' length & 4.5 & m\\\hline
			Slide mass & 9 & kg\\\hline
			Spring maximum compression & 0.32 & m\\\hline
			Spring stiffness & 60 &N/m\\\hline
			Battery bank capacity & 260 &Ah\\\hline
			Battery bank voltage & 12 &V\\\hline
			Inverter continuous power & 3000 & W\\\hline
			Inverter peak power & 6000&W\\\hline\hline
			\multicolumn{3}{|l|}{\textbf{Glider}}\\\hline\hline
			Wingspan & 1.68 &m\\\hline
			Mean Aerodynamic Chord & 0.194 &m\\\hline
			Aspect Ratio& 8.89&-\\\hline
			Wing Surface Area& 0.3174 & m$^2$\\\hline
			Mass &  1.2 &kg\\\hline
			Peak Motor Power &360 &W\\\hline
			Rated Motor Power &290 &W\\\hline
			Battery voltage &14.4 &V\\\hline
			Propeller diameter & 0.22&m\\\hline
		\end{tabular}
	\end{table}

	\subsection{Aircraft}\label{SS:Glider}
	
	We consider a rigid aircraft with conventional design, equipped with an horizontal axis propeller and with four control surfaces - elevator, ailerons, rudder and flaps. As a matter of fact, in our control approach we use only the motor, ailerons and elevator to control the aircraft, since these proved to be sufficient to achieve good performance. 
	Further improvements can be obtained by exploiting also the remaining control surfaces. A mechanism to attach and detach the tether is also installed on the aircraft. The following onboard measurements are available to be used for feedback control: position and velocity with respect to the ground; roll, pitch and yaw angles and their rates; 3D accelerations with respect to a local (i.e. fixed with the aircraft) reference frame; airspeed along the local longitudinal axis.
	
\begin{figure}[h]
		\centerline{ \includegraphics[width=09cm,clip]{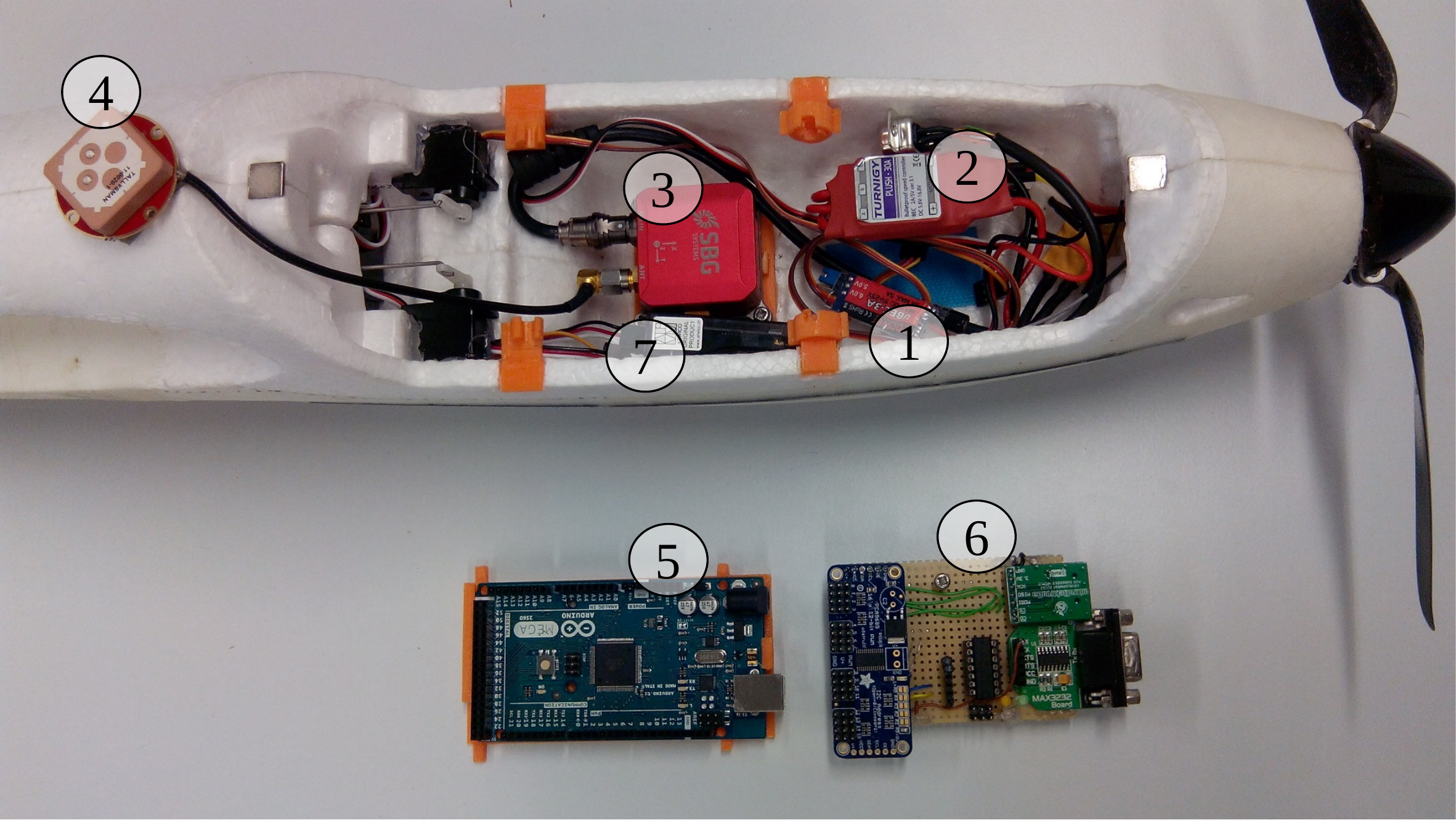}}
		\caption{Onboard measurement and control hardware for the glider. 1. DC-DC power converter; 2. motor driver; 3. inertial measurement unit; 4. GPS antenna; 5. Arduino$^\circledR$ Mega 2560 embedded platform; 6. extension boards; 7. RC receiver.}\label{P:Disassembled}
	\end{figure}

	In our prototype, we use a commercial model glider made of styrofoam (visible in Fig. \ref{F:proto_pic}), which we reinforced with fiberglass and modified to add the tether attachment mechanism. In the glider we installed an inertial measurement unit (IMU) (an SBG Systems ``Ellipse-N'' unit \cite{SBGSystems2016}) and a pitot probe, which together provide the above-mentioned measurements at a rate of 50 Hz, large enough for the considered control problem since the aircraft dynamics exhibit dominant poles lower than 5 Hz (see sections \ref{S:design}-\ref{S:results} for more details on the time constants of the glider dynamics). We also installed an embedded platform (an Arduino MEGA 2560 board \cite{Arduino2016}) to fuse all the measurements, log the data, and compute and actuate the control inputs. A commercial remote controller and the corresponding onboard receiver are used to pilot the aircraft manually if needed, and to switch between manual and automatic flight modes. Fig. \ref{P:Disassembled} shows the inside of the glider's fuselage, with the IMU and the control board. The control surfaces and the tether release mechanism are actuated by standard servomotors widely employed in the model glider business, and the front propeller is driven by a standard brushless DC motor. For the sake of brevity, only the main characteristics of the glider are shown in Table \ref{T:gs_data}, while the full details on the employed hardware, sensors, actuators etc. are reported in \cite{Fagiano2016submitted}.
	
	\subsection{Ground station control}\label{SS:GS_control}
	As anticipated in the introduction, the details of the ground station design and control are outside the scope of this paper and can be found in \cite{Fagiano2016submitted}. However, we recall here the principle of operation of the winch control system, since this is instrumental to understand one of the main assumptions that we consider when modeling the aircraft and designing its onboard controller.
	
	A first goal for the winch control strategy is to limit the force exerted by the tether on the aircraft to relatively small values. This aspect is fundamentally different from what is usually considered in AWE during the power generation phase, where the tether force is very large. The main reason for this difference is that, during power generation, crosswind trajectories are flown at large speed and the aircraft's velocity vector is roughly perpendicular to the tether, thus the tether force does not affect the aircraft forward velocity. On the contrary, during take-off the aircraft speed is much lower and its velocity vector is almost aligned with the tether, such that even a small tether force can slow-down the aircraft and provoke a stall situation.
	
	At the same time, reeling out with no load must be also avoided, in order to prevent entanglement of the tether and to limit the tether sag, which increases the weight and drag forces acting on the aircraft and might as well lead to entanglement with obstacles on the ground.
	Such conflicting control objectives (low tether force and low tether sag) require some sort of coordination between the aircraft and the ground station. A natural approach to this problem would be to exploit communication between the two sub-systems to try to coordinate their motion. However, such an exchange of information between the controllers might be still not enough to achieve the required coordination, since one would need to gather not only the aircraft velocity vector and the tether length and speed, but also the relative angle between the aircraft velocity and the tether. The latter might be very difficult to estimate or measure with good enough accuracy. Furthermore, even with perfect knowledge of the position and velocity of both the aircraft and tether/winch, the torque limitation and the inertia of the winch might still prevent the control system from acting fast enough to avoid a stall situation when the tether force becomes too large.
	
\begin{figure}[!h]
		\begin{center} 
			\includegraphics[width=12cm,clip]{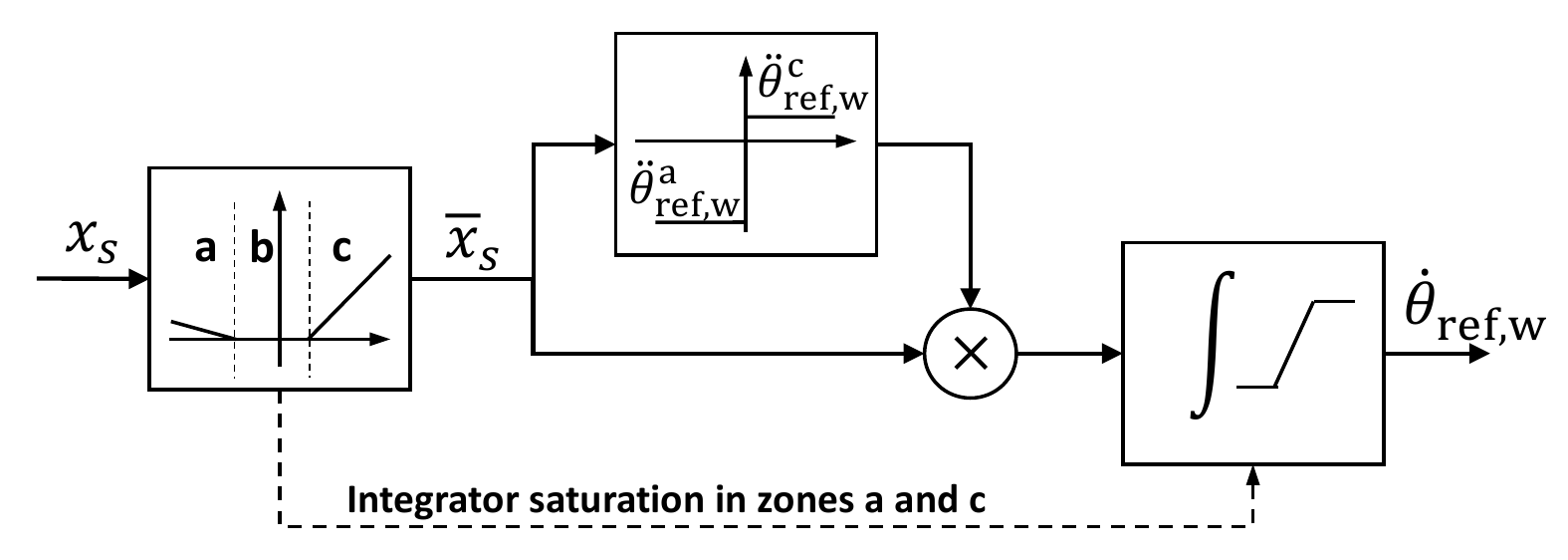}
			\caption{Block-diagram of the feedback contribution to the reference winch speed.}\label{F:winch_control_sketch}
		\end{center}
	\end{figure}

	In our study, a key design choice we made is to have no exchange of information between the ground station controller and the aircraft. The main advantage of this choice is that communication reliability issues and delays do not have to be dealt with, while the main disadvantage is that the achievable system performance, for example in terms of flight pattern repeatability and tolerance to wind disturbances, is inferior to the case when communication between the aircraft and the ground is exploited. Since our activity was conceived as a proof of concept and performance optimization was not a priority, we decided not to exploit communication. In order to solve the winch/aircraft coordination problem without using an active exchange of information, we designed the above-mentioned mass-spring system. The employed spring is relatively soft, such that a small pulling force on the tether produces a rather large displacement. The spring compression helps to reduce the harshness with which the tether pulls on the aircraft, and gives more time to the winch control system to react. The spring displacement is measured by a potentiometer and employed as a feedback variable to compute the reference reeling speed for the winch. A standard proportional  controller in an inner control loop is then in charge of tracking such a reference speed by adjusting the motor current, which is subject to saturation (hence imposing a saturation on the motor torque). The full details of the ground station controller and more in general of the ground station design are described in \cite{Fagiano2016submitted}. However, for the sake of completeness we include here a short description of the feedback control strategy that computes the reference winch speed.\\
	More specifically, let us denote with  $x_\text{s}(k)$ the measured spring compression, where $k\in\mathbb{Z}$ indicates the discrete time instants employed by the ground station controller. We set two threshold values, $x_\text{s}^I,\,x_\text{s}^{II}$, which divide the available spring travel in three zones:
	\begin{itemize}
		\item \textbf{Zone a} ($0\leq x_\text{s}(k)\leq x_\text{s}^I$): the spring is practically uncompressed, the winch shall decrease speed and eventually reel-in;
		\item \textbf{Zone b} ($x_\text{s}^I < x_\text{s}(k)<x_\text{s}^{II}$): the spring is subject to low force, the winch speed shall be held (constant tether speed);
		\item \textbf{Zone c}: ($x_\text{s}^{II}\leq x_\text{s}(k) \leq \overline{x}_\text{s} $): the spring is subject to relatively large force, the winch shall increase its speed and reel-out to release the tether.
	\end{itemize}
	Then, the reference winch speed, $\dot{\theta}_\text{ref,w}(k)$, is computed according to the following strategy (see Fig. \ref{F:winch_control_sketch}): \small
	\begin{equation}\label{E:fbck_winch}\nonumber
	\begin{array}{l}
	\verb"If " 0\leq x_\text{s}(k)<x_\text{s}^I \text{ (\textbf{Zone a})}\\
	\\
	\;\;\bar{x}_\text{s}(k)=\dfrac{x_\text{s}(k)-x_\text{s}^I}{x_\text{s}^{I,\text{a}}-x_\text{s}^I}\\
	\;\;\dot{\theta}_\text{ref,w}(k)=\min\left(0,\max\left(\underline{\dot{\theta}}_\text{ref,w}^\text{fbck},\,
	\left(\dot{\theta}_\text{ref,w}(k-1)+T_s\ddot{\theta}_\text{ref,w}^\text{a}\bar{x}_\text{s}(k)\right)\right)\right)\\
	\\
	\verb"Else if " x_\text{s}^I \leq x_\text{s}(k)<x_\text{s}^{II} \text{ (\textbf{Zone b})}\\
	\\
	\;\;\dot{\theta}_\text{ref,w}(k)=\dot{\theta}_\text{ref,w}(k-1)\\
	\\
	\verb"Else if " x_\text{s}^{II}\leq x_\text{s}(k) \leq \overline{x}_\text{s} \text{ (\textbf{Zone c})}\\
	\\
	\;\;\bar{x}_\text{s}(k)=\dfrac{x_\text{s}(k)-x_\text{s}^{II}}{x_\text{s}^{II,\text{c}}-x_\text{s}^{II}}\\
	\;\;\dot{\theta}_\text{ref,w}(k)=\max\left(0,\min\left(\overline{\dot{\theta}}_\text{ref,w}^\text{fbck},\,
	\left(\dot{\theta}_\text{ref,w}(k-1)+T_s\ddot{\theta}_\text{ref,w}^\text{c}\bar{x}_\text{s}(k)\right)\right)\right)\\
	\end{array}
	\end{equation}\normalsize
	where $\underline{\dot{\theta}}_\text{ref,w}^\text{fbck},\,\overline{\dot{\theta}}_\text{ref,w}^\text{fbck}$ are the desired minimum and maximum reference speed values that can be issued, and $\ddot{\theta}_\text{ref,w}^\text{a},\,\ddot{\theta}_\text{ref,w}^\text{c}$ are the desired angular accelerations for the reference speed. Such values are scaled according to the position of the potentiometer relative to the values $x_\text{s}^{I,\text{a}}<x_\text{s}^I$ and $x_\text{s}^{II,\text{c}}>x_\text{s}^{II}$, respectively for zones \textbf{a} and \textbf{c}, which are design parameters as well. Equation \eqref{E:fbck_winch} represents an integral controller, where the integrated quantity is the distance of the spring position $x_\text{s}(k)$ from the interval $(x_\text{s}^I,\,x_\text{s}^{II})$ (i.e. zone \textbf{b}) and the gain is piecewise constant, since it is different in zones \textbf{a} and \textbf{c}. Moreover, a saturation of the integrated variable to negative (resp. positive) values is operated whenever the spring enters zone \textbf{a} (resp. \textbf{c}), in order to quickly start to reel-in (resp. reel-out) when the tether is released (resp. pulled).  As shown in \cite{Fagiano2016submitted}, a sensible choice for the involved design parameters is $\underline{\dot{\theta}}_\text{ref,w}^\text{fbck},\,\ddot{\theta}_\text{ref,w}^\text{a}<0$, $\overline{\dot{\theta}}_\text{ref,w}^\text{fbck},\,\ddot{\theta}_\text{ref,w}^\text{c}>0$, $x_\text{s}^{I,a}\approx x_\text{s}^I/2$ and $x_\text{s}^{II,c}\approx (\overline{x}_\text{s}+x_\text{s}^{II})/2$.
	
	The actual winch control strategy is slightly more complex, since it involves also a feed-forward term during the take-off transient, in order to latch the winch speed to the slide speed, see \cite{Fagiano2016submitted} for details.
	
\begin{figure}[!tbh]
		\begin{center}
			\includegraphics[width=9cm]{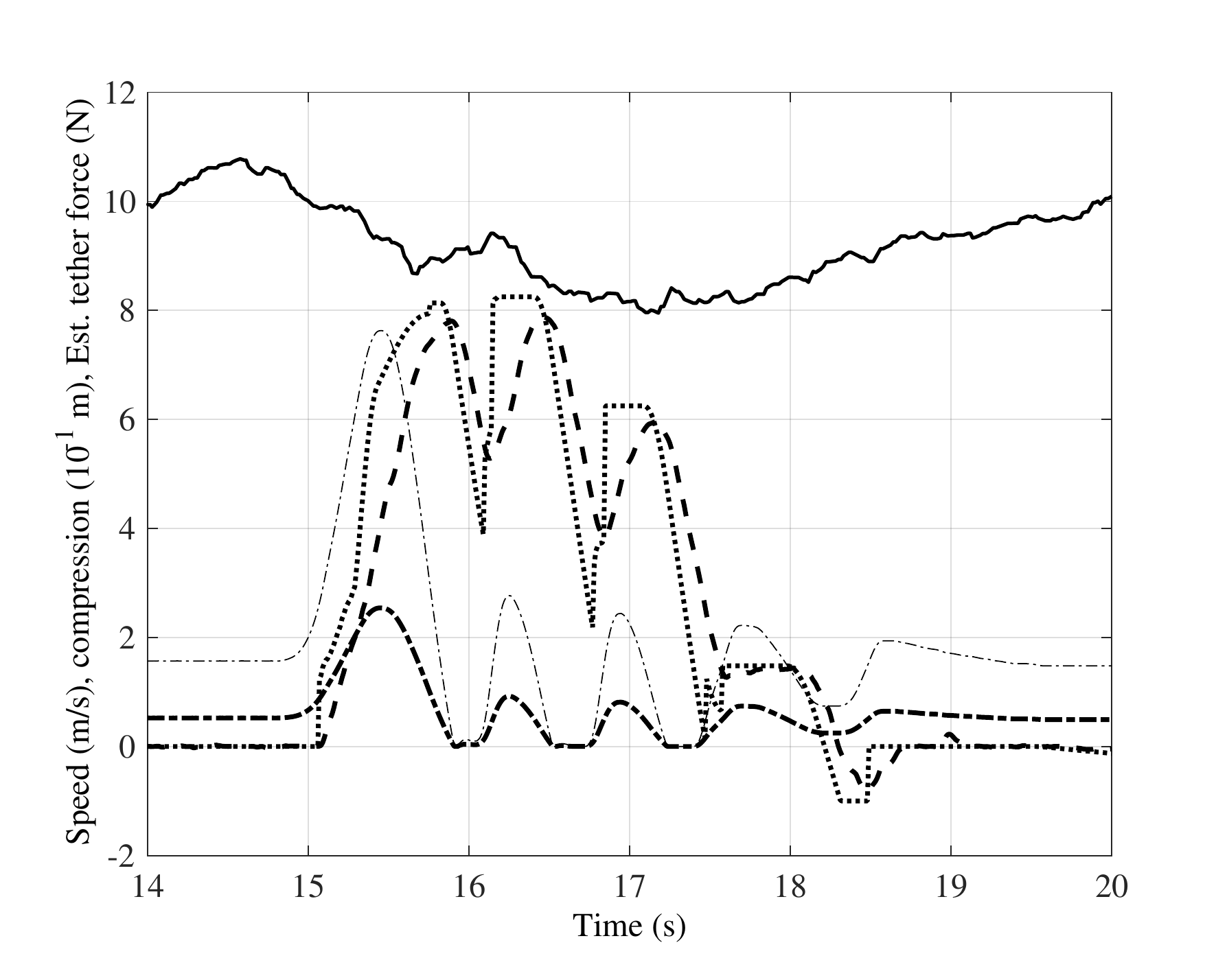}
			\caption{Experimental results: example of transient of spring compression and related loss of speed of the aircraft when the tether is taut. Solid line: airspeed (m/s), dashed: winch speed (m/s), dotted: reference winch speed (m/s), dash-dot: spring compression (10$^{-1}\,$m). Thin dash-dotted line: tether force estimated from the spring compression.}
			\label{F:winch_glider_exp_example}
		\end{center}
	\end{figure}

	An example of the typical behavior obtained with the described winch control system is shown in Fig. \ref{F:winch_glider_exp_example}, which presents the aircraft airspeed, winch speed,  reference winch speed, spring compression and an estimate of the tether force during autonomous tethered flight, while the aircraft is pulling on the tether and then starting a turn maneuver. The difference between the aircraft airspeed and the tether speed is due to the relative orientation between the two. Referring to Fig. \ref{F:winch_glider_exp_example}, the initial spring position is such that, according to the winch control strategy, the reeling speed is held constant (equal to zero in this specific case). Then, at $t\simeq15\,$s the tether becomes taut and the spring compresses, entering a region where the reference winch speed is increased (according to the winch control law). When the tether length becomes larger than the distance between the aircraft and the ground station, the tether becomes slack again and the spring is released. Then, the reference winch speed is first held in place (as the spring travels through the "hold" region) and then quickly ramped down toward negative values. However, as the winch slows down, the aircraft might pull again on the tether, hence giving place to the oscillating behavior of the winch speed, until finally after $t\simeq18.5\,$s the tether force settles again on a roughly constant value, corresponding to a fixed spring position. This kind of transient is very common with the adopted strategy, where the winch controller and the aircraft controller are fully decoupled and coordination has to be done via the spring compression. An oscillating behavior of the spring while the tether is taut indicates that the (average) reel-out speed is matching well with the aircraft speed projected along the tether: the alternative would be either to reel-out too fast, with consequent full decompression of the spring and tether entanglement on the ground, or to reel-out too slowly, with consequent stall of the aircraft.
	
	It can be noted that the typical perturbation of the glider speed due to the tether force is less than 1 m/s, which is an acceptable performance for the sake of our study. A rough estimate of the the tether force, neglecting the inertia and viscous friction of the spring mechanism, can be obtained by simply scaling the spring compression by its stiffness, of about 60 N/m, and considering that, since the tether runs 180 deg around the pulley attached to the spring (see Fig. \ref{F:sys_pic}),  the force acting on the spring equals two times the tether force. Such an estimate is plotted in Fig. \ref{F:winch_glider_exp_example}, too, showing that impulses of about 3-8 N peak and 0.3-0.7 s of duration are experienced by the aircraft. These values are acceptable for the glider used in our tests, since they don't compromise the flight dynamics.
	
	The described winch control strategy and its performance represent the main motivation for which, in the remainder of this paper, we will refrain from including a model of the tether or of the ground station to design the aircraft controller. In fact, we will consider a rather simple model of the aircraft dynamics, where the force exerted by the tether is embedded, together with other neglected aspects and high-order dynamics, in an unknown-but-bounded additive disturbance term.
	
	
	\subsection{System operation and control objectives}\label{SS:objectives}
	
	The overall aim of the onboard controller is to reach a repetitive flight pattern above the ground station, starting from a standstill position on the slide. We divided such a control task in two subsequent phases, whose control objectives are described next.
	
	\textbf{Take-Off and climbing:} The glider is initially attached to the slide. The take-off maneuver is initiated by the ground station, which accelerates the slide. The onboard controller has to detect this situation (we recall that there is no exchange of information between the ground station controller and the onboard one) and stabilize the aircraft's attitude until it detaches from the slide, while at the same time powering up the onboard motor to sustain the climb. After detaching from the slide, the first goal of the controller is to reach a safe altitude, before performing any other maneuver. A relatively large climb angle is thus targeted to rapidly reach such a safe altitude. A large climbing angle is also beneficial when the tether is taut, since a smaller component of the tether force adds to the aerodynamic drag.
	
	
	\textbf{Transition phase and figure-of-eight patterns:}
	Once a safe altitude has been reached, the controller shall carry out a  transition maneuver to engage a repetitive figure-of-eight pattern parallel to the ground at a target altitude, withstanding the presence of wind gusts and of the tether. Assuming that the wind flow is mainly parallel to the ground as well, such a pattern gives rise to low tether forces. In a power generating system, this phase can be then exploited to check the system and estimate the wind conditions before carrying out a further transition  to crosswind figure-of-eight paths. Such a transition can be achieved by smoothly adjusting the eight-pattern from horizontal to crosswind. 
	At the same time, the winch control system has to be changed from a ``low-tether-force'' strategy, like the one considered in this paper (which allows the aircraft to take-off notwithstanding the tether), to a ``high-tether-force'' strategy, e.g. where the reel-in torque is set proportional to the square of the reel-out speed, see e.g. \cite{Zgraggen2016}. Since in our study the goal is to demonstrate only the autonomous take-off and the transition to low-force flight, we stop the experiments once the horizontal eight-pattern has been reached. 

	\section{System modeling and control design}\label{S:design}
	
	We will now describe in detail the design of a controller able to achieve the objectives described in section \ref{SS:objectives}. The design is model based: we first derive a relatively simple model of the aircraft dynamics and identify its parameters using experimental data collected in preliminary flights, and then we design the final controller on the basis of this model.
	
	\subsection{Aircraft modeling and parameter identification}\label{SS:model_id}
	
	\subsubsection{Model equations}\label{SSS:model}
	There exist a vast literature and many textbooks concerned with the flight mechanics of a rigid aircraft like the one considered in this work, albeit without the tether. A frequent approach is to employ a six-degrees-of-freedom (dof) model with lumped parameters accounting for the geometrical, inertial and aerodynamic characteristics of the aircraft. In a preliminary work we used such a model (taken from \cite{etkin}), together with models of the ground station and of the tether, in order to obtain a simulation tool for our control algorithms \cite{Nguyen-Van2016}. This model is quite realistic as it includes the most important system nonlinearities, e.g. the stall behavior, and considers the coupling among the different states. However, it is rather complex, since it comprises 12 states (three-dimensional position and rotation and their time derivatives) and 5 inputs (motor, rudder, ailerons, elevator, and flaps). If the presence of the ground station (slide, winch and mass-spring system) and of the tether were also considered (the latter with a simple purely elastic model as done in \cite{Nguyen-Van2016}), the number of states and inputs would raise to  18 and 7, respectively. Such a model is very useful to simulate the system and carry out a preliminary test of control algorithms, however its use for control design is not trivial. In fact, even the parameter identification phase raises difficulties 
	due to the presence of nonlinearities, open-loop unstable behavior and poorly controllable and expensive experimental tests, where the wind vector, which would be required for an accurate identification of the aerodynamic characteristics, is anyways not measured.
	
	Therefore, for the sake of control design we decided to take a radically opposite approach and adopt a very simple model of the system dynamics to attain our goals, with the same spirit of previous contributions on AWE control design, \cite{Erhard2012a,Fagiano2014,Erhard2015}. Namely, we consider two main dynamical modes: the first one to describe the turning behavior, and the second one to describe the vertical motion.  A third dynamical equation describes the link between the airspeed and the propeller thrust. The couplings among all these modes are neglected and embedded into additive disturbance/uncertainty terms. The core building blocks of the resulting model are either linear time-invariant, second-order dynamical equations whose parameters can be easily identified from the available measurements, or kinematic relationships that feature no parameter at all. Thus, the parameter identification phase is rather straightforward and effective, as shown below. Moreover, the model is readily usable for control design using a hierarchical approach, as detailed in section \ref{SS:Autopilot_control}.
	
	We start by introducing an inertial coordinate system $(X,Y,Z)$ fixed to the ground with the $Z-$axis pointing up, and a local one $(x,y,z)$ fixed to the aircraft (body frame), with the $z-$axis pointing down when the aircraft is in horizontal flight, see Fig. \ref{F:sys_pic}. The rails' orientation (take-off direction) forms an angle $\gamma_\text{ref,to}$ with the inertial $X$-axis. The position vector of the aircraft is denoted by $\boldsymbol{p}$ (bold letters indicate vectors in a three-dimensional space), its velocity by $\dot{\boldsymbol{p}}\doteq d\boldsymbol{p}/dt$, where $t$ is the continuous time variable. When considering the components of a vector, we add as a subscript the corresponding axis, e.g. $p_X$ is the component of vector $\boldsymbol{p}$ along the inertial $X-$axis. We further denote by $\varphi,\theta,\psi$ the standard roll, pitch and yaw angles (Euler angles), 
	and with $v_a$ the magnitude of the airspeed toward the body $x-$axis direction. The full airspeed vector is in principle computed as $\boldsymbol{W}-\dot{\boldsymbol{p}}$, where $\boldsymbol{W}$ is the wind velocity vector, which however is not measured in practice. On the other hand, the value of $v_a$ is measured thanks to the onboard pitot probe aligned with the body $x-$axis. Finally, we denote with $u_\varphi$, $u_\theta$ and $u_{\text{m}}$ the control inputs for the ailerons, elevator, and propeller thrust, respectively.
	
	The equations for the first mode (turning behavior) are derived starting from the roll angle dynamics:
	\begin{equation}\label{E:roll}
	\ddot{\varphi}(t)=a_{\varphi}\dot{\varphi}(t)+b_{\varphi}u_{\varphi}(t)+d_{\varphi}(t)\\
	\end{equation}
	where $a_{\varphi}$ and $b_{\varphi}$ are parameters to be identified, and $d_{\varphi}$ is a disturbance term accounting for neglected dynamics, wind, and the presence of the tether. The parameter $a_{\varphi}$ is expected to be negative, so to have a stable first-order model linking the aileron input $u_{\varphi}$ to the roll rate $\dot{\varphi}.$ Equation \eqref{E:roll} is motivated by physical insight, since a constant aileron input gives rise to a constant roll rate in steady-state(due to the equilibrium of the aerodynamic roll moment induced by the aileron with the opposite one generated by the roll rate), by experimental evidence (see section \ref{SSS:param_id} below) and by the structure of the equation obtained by linearizing a more complex model, like the one derived in \cite{Nguyen-Van2016}, in a steady-state, straight-flight configuration.
	
\begin{figure}[!hbt]
		\centerline{		\includegraphics[width=9cm,clip]{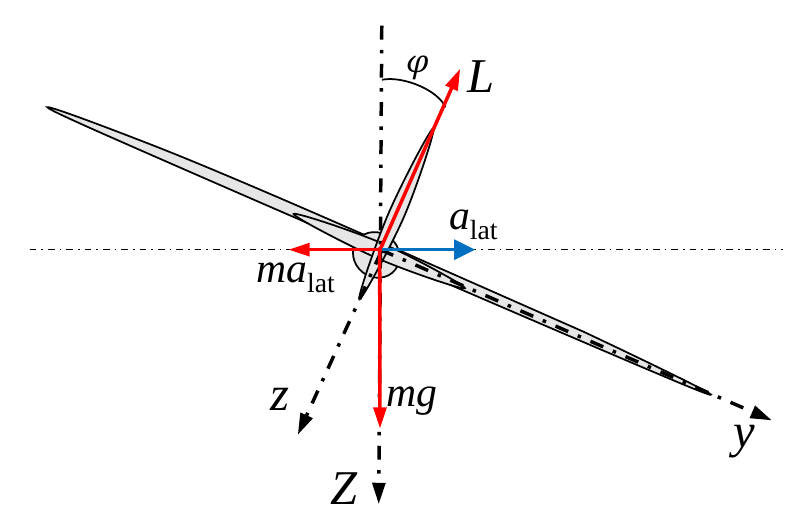}}
		\caption{Sketch of the considered turning model for the aircraft}\label{F:RollAngle}
	\end{figure}

	Considering a turning maneuver at constant tangential speed and with constant radius along a circular trajectory parallel to the ground (i.e. the $(X,Y)$-plane), the roll angle $\varphi$ can be then linked to the lateral acceleration of the aircraft by considering the equilibrium of aerodynamic lift, gravitational force, and centrifugal force during the turn:
	\begin{subequations}\label{E:lift_equilibrium}
		\begin{align}
		L\sin{(\varphi)}=m a_\text{lat}\label{E:lateral_lift_eq}\\
		L\cos{(\varphi)}=m g\label{E:vertical_lift_eq}
		\end{align}
	\end{subequations}
	where $L$ is the aerodynamic lift force, $m$ the mass of the aircraft, and $a_\text{lat}$ is the centripetal acceleration. Fig. \ref{F:RollAngle} presents an intuitive representation of equations \eqref{E:lift_equilibrium}.
	As mentioned, such equations hold with good approximation when the path is parallel to the ground, such that when $\varphi=0$ the lift force points along the inertial $Z$ direction. Moreover, under the considered assumptions the tangential speed equals $|\dot{\boldsymbol{p}}(t)|$ and from the kinematics of rigid bodies we have:
	\begin{equation}\label{E:lateral_acc}
	a_\text{lat}=|\dot{\boldsymbol{p}}(t)|\dot{\psi}
	\end{equation}
	where $\dot{\psi}$ is the yaw rate. Now, inserting \eqref{E:lateral_acc} into \eqref{E:lateral_lift_eq}, dividing \eqref{E:lateral_lift_eq} by \eqref{E:vertical_lift_eq} and assuming small roll angles, such that we can use the linearization of the trigonometric functions at $\varphi=0$ (in our experiments we seldom exceeded 30$^\circ$ of roll), we obtain:
	\begin{equation}\label{E:yaw_rate}
	\dot{\psi}(t)=\dfrac{g}{|\dot{\boldsymbol{p}}(t)|}\varphi(t)
	\end{equation}
	Finally, it is useful to also introduce the link between the trajectory curvature $1/R(t)$ (where $R(t)$ is the turning radius) and the roll angle in the model. Considering that $\dot{\psi}(t)\simeq\dfrac{|\dot{\boldsymbol{p}}(t)|}{R(t)}$, we have:
	\begin{equation}\label{E:curvature}
	\dfrac{1}{R(t)}=\dfrac{g}{|\dot{\boldsymbol{p}}(t)|^2}\varphi(t)
	\end{equation}
	Together, equations \eqref{E:roll} and \eqref{E:yaw_rate} form our model for the turning behavior, since they link the aileron input to the roll angle, and the latter to the flown trajectory on a plane parallel to the ground. Such equations have the advantage of having just two parameters to be identified within a second-order, linear time invariant (LTI) system (eq. \eqref{E:roll}), while all the other involved quantities are measured by the available sensors, in particular magnetometers for $\varphi$, gyroscopes for $\dot{\psi}$, GPS and accelerometers for $|\dot{\boldsymbol{p}}(t)|$.
	
\begin{figure}[!hbt]
		\centerline{		\includegraphics[width=9cm]{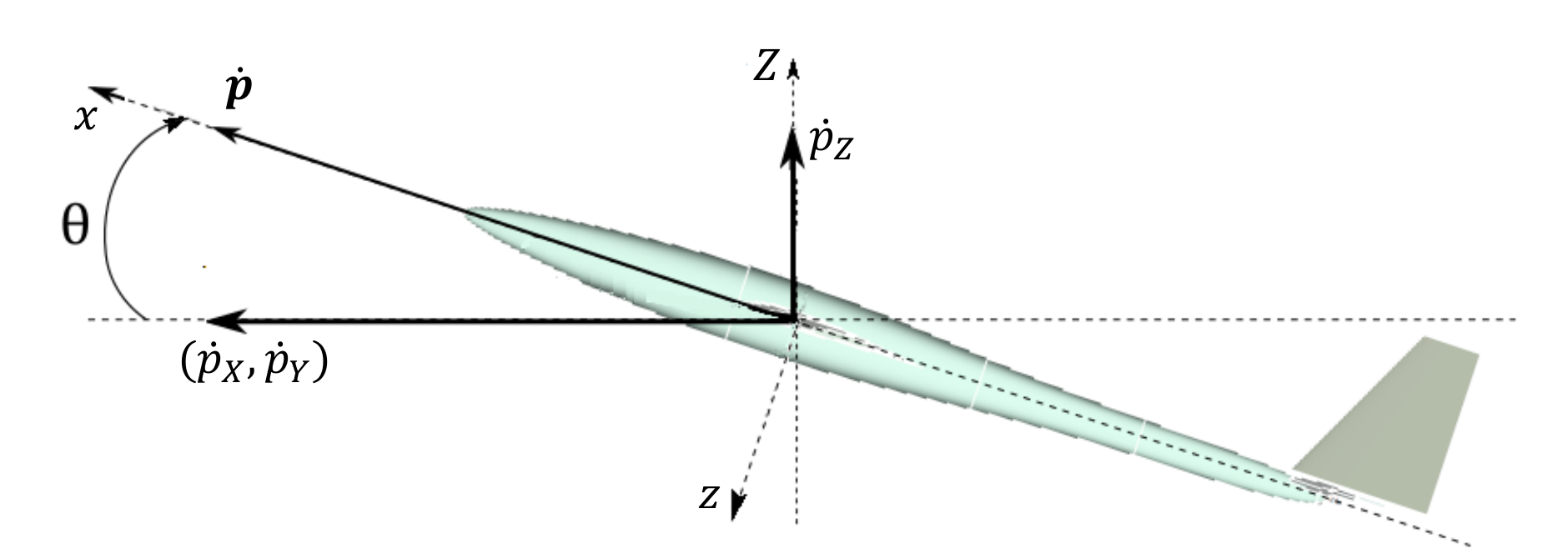}}
		\caption{Sketch of the considered pitch/vertical motion model for the aircraft}\label{F:PitchAngle}
	\end{figure}

	About the vertical motion, we start from the pitch dynamics:
	\begin{equation}\label{E:pitch}
	\ddot{\theta}(t)=a_{\theta}\dot{\theta}(t)+b_{\theta}u_{\theta}(t)+d_{\theta}(t)
	\end{equation}
	where, in a way similar to equation \eqref{E:roll}, $a_{\theta}$ and $b_{\theta}$ are unknown parameters to be identified and $d_{\theta}$ accounts for  neglected effects. Again, the parameter $a_{\theta}$ is expected to be negative, i.e. the dynamical behavior from  the elevator input $u_{\theta}$ to the pitch rate is asymptotically stable. The model \eqref{E:pitch} is also based on physical insight and prior knowledge, it is confirmed by the experimental data and it holds within certain limits, in particular for small values of the pitch angle and of the angle of attack. 
	Such conditions are met in our experiments. Assuming now a straight-line flight parallel to the ground at constant forward speed, a constant pitch value $\theta_0$ exists, such that this motion is at steady state, i.e. the lift force equals the weight of the aircraft and the propeller's thrust counteracts the drag. For simplicity and without loss of generality, we assume $\theta_0=0$. Exploiting again the assumption of small $\theta$ values, we can then approximate the link between the pitch angle and the vertical velocity along the inertial $Z-$axis as:
	\begin{equation}\label{E:vertical speed}
	\dot{p}_Z(t)=|\dot{\boldsymbol{p}}(t)|\theta(t)
	\end{equation}
	Fig. \ref{F:PitchAngle} provides a graphical interpretation of \eqref{E:vertical speed}.
	Equations \eqref{E:pitch} and \eqref{E:vertical speed} link the pitch dynamics to the vertical inertial component of the aircraft trajectory, and form our model of the vertical dynamics. Also in this case there are 
	only few parameters, which can be easily identified from experimental data. 
	
	Last, we employ the following equations to model the effect of the propeller thrust on the airspeed $v_a$:
	\begin{equation}\label{E:airspeed}
	u_{\text{m}}(t)=\dfrac{1}{2}\rho A C_D v_a(t)^2+d_{v_a}(t)
	\end{equation}
	where $\rho$ is the air density, $C_D$ the aerodynamic drag coefficient, $A$ the effective area of the aircraft, and $d_{v_a}(t)$ a term accounting for neglected effects. Equation \eqref{E:airspeed} is derived by assuming equilibrium between the motor thrust and the aerodynamic drag, and a small angle of attack (i.e. assuming that the apparent wind is mainly aligned with the body $x-$axis). Similarly to the previous equations of motion, we include the approximation errors arising from such an assumption in the term $d_{v_a}(t)$. The simple equation \eqref{E:airspeed} is good enough for the considered application. 
	
	Equations \eqref{E:roll}, \eqref{E:yaw_rate}, \eqref{E:pitch}, \eqref{E:vertical speed} and \eqref{E:airspeed} form the model of the aircraft that we will use for control design. The only unknown parameters are $a_\varphi,\,b_\varphi,\,a_\theta,\,b_\theta$, which can be identified from data as shown in the following. Regarding the value of $C_D$ in \eqref{E:airspeed}, a good estimate can be obtained from the airfoil shape and the aircraft design; moreover such a parameter is not critical for  control design as we comment later on in section \ref{SSS:tuning}.
	
	\subsubsection{Parameter identification}\label{SSS:param_id}
	
\begin{figure}[!h]
		\centerline{		\includegraphics[width=9cm,clip]{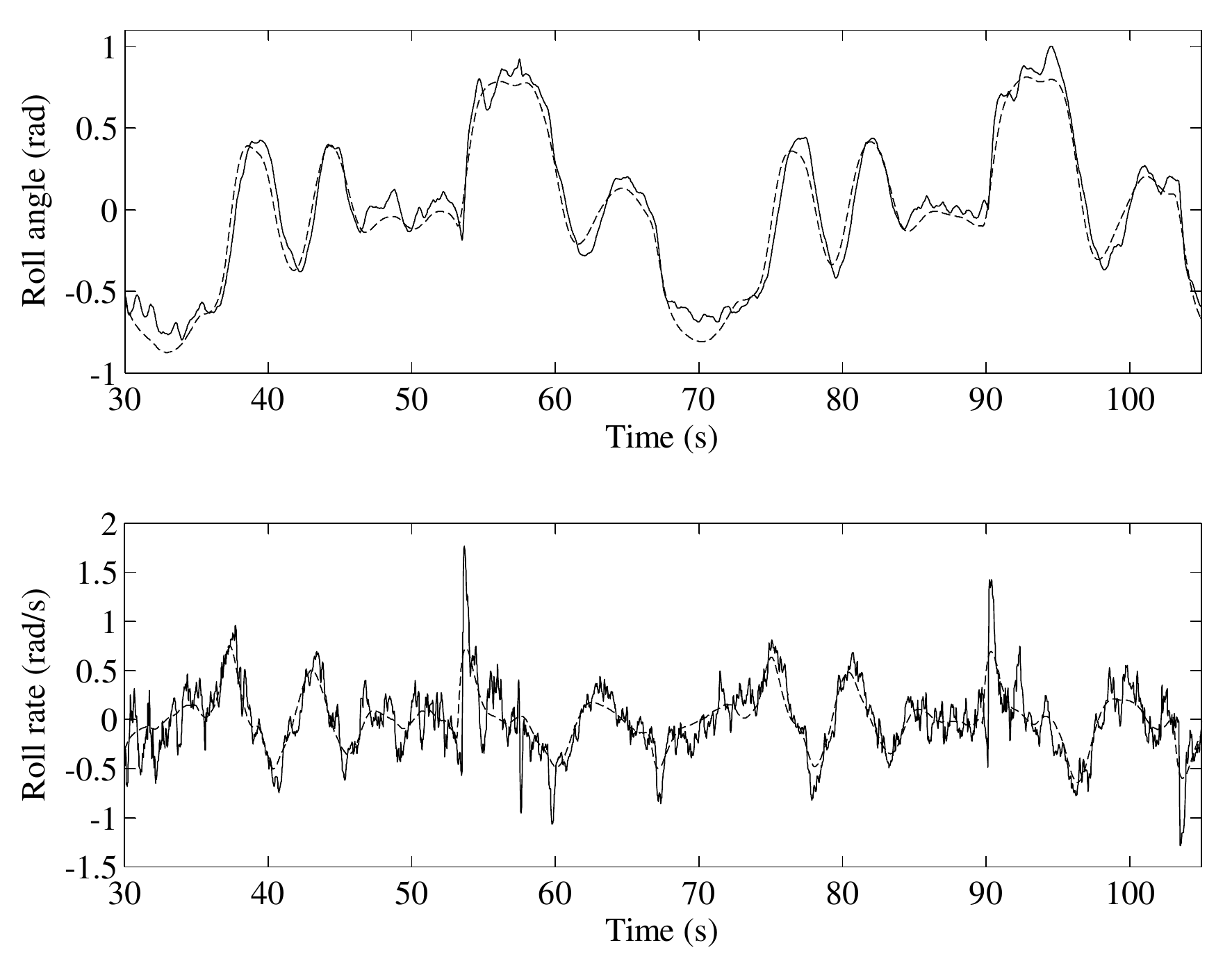}}
		\caption{Experimental results. Solid lines: measured roll angle (upper plot) and roll rate (lower plot) obtained in fly-by-wire mode with a preliminary controller. Dashed lines: roll angle and rate obtained with the model \eqref{E:roll}, whose parameters have been identified by solving problem \eqref{E:opt_id}. The data shown in the plot are different from those used for the identification.}\label{G:p_curve_fit}
	\end{figure}

	For the sake of brevity, we present here the identification approach and results only for the roll angle dynamics \eqref{E:roll}. For the pitch angle, we employed the same approach and obtained similar results. Equation (\ref{E:roll}) involves two parameters to identify, $a_\varphi$ and $b_\varphi$. We first estimated a value for both parameters by linearizing the 6-dof model of \cite{Nguyen-Van2016} and designed a preliminary proportional controller $K_\text{id}$ to track a reference roll angle $\varphi_{\text{ref}}$, and we tested such a controller in fly-by-wire, i.e. with a human pilot issuing the desired roll angle from the ground. Using a batch of the collected data, we then identified the model parameters in closed-loop with the same preliminary controller by solving the following optimization problem:
	\begin{subequations}\label{E:opt_id}
		\begin{gather}
		\min_{a_\varphi,b_\varphi} \left\|\{\tilde{\varphi}\}_1^N-\{\varphi\}_1^N\right\|_2 +\left\|\{\tilde{\dot{\varphi}}\}_1^N-\{\dot{\varphi}\}_1^N\right\|_2 \\
		\textrm{subject to}   \nonumber\\
		\dot{\varphi}(\tau+1)= \dot{\varphi}(\tau)+T_s \left(a_\varphi \dot{\varphi}(\tau) + b_\varphi K_\text{id} (\tilde{\varphi}_{\text{ref}}(\tau)-\varphi(\tau))\right)\\
		\varphi(0)=\tilde{\varphi}(0)\\
		\dot{\varphi}(0)=\tilde{\dot{\varphi}}(0)
		\end{gather}
	\end{subequations}
	where $T_s$ is the employed sampling time, $\tau=0,1,\ldots,N-1$ denotes the discrete-time sampling instants, $N$ is the number of employed samples, $\{\tilde{\varphi}\}_1^N\doteq[\tilde{\varphi}(1),\ldots,\tilde{\varphi}(N)]^T$ is the sequence of measured roll angle values and $\{\varphi\}_1^N$ of estimated ones (and similarly $\{\tilde{\dot{\varphi}}\}_1^N,\,\{\dot{\varphi}\}_1^N$ are the measured and estimated roll rate values). The notation $\tilde{}$ in \eqref{E:opt_id} denotes the collected noise-corrupted experimental data. Namely, problem \eqref{E:opt_id} aims to find values of the parameters that minimize the 2-norm of the error between the measured and simulated roll rate, in the presence of the same preliminary controller and applying the measured input reference roll angle.
	The resulting problem is a nonlinear program (NLP) which we solved for a local minimum by using a sequential quadratic programming approach (the function \verb"fmincon" in Matlab$^\circledR$). An example of the obtained results is illustrated in Fig~\ref{G:p_curve_fit}, where the roll rate obtained using the parameters computed by solving \eqref{E:opt_id} is compared with the measured one, using a second batch of data not employed for the parameter identification. The obtained results further confirm  that the considered second order model \eqref{E:roll} approximates the roll motion well and that we can avoid the complexity of the 6-dof model for the controller design. In section \ref{SS:implementation}, we present the obtained numerical values of the parameters. 
	
	\subsection{Control Design}\label{SS:Autopilot_control}
	
	\subsubsection{Controller structure}\label{SSS:structure}
\begin{figure}[hbt]
		\centerline{
			\includegraphics[width=16cm]{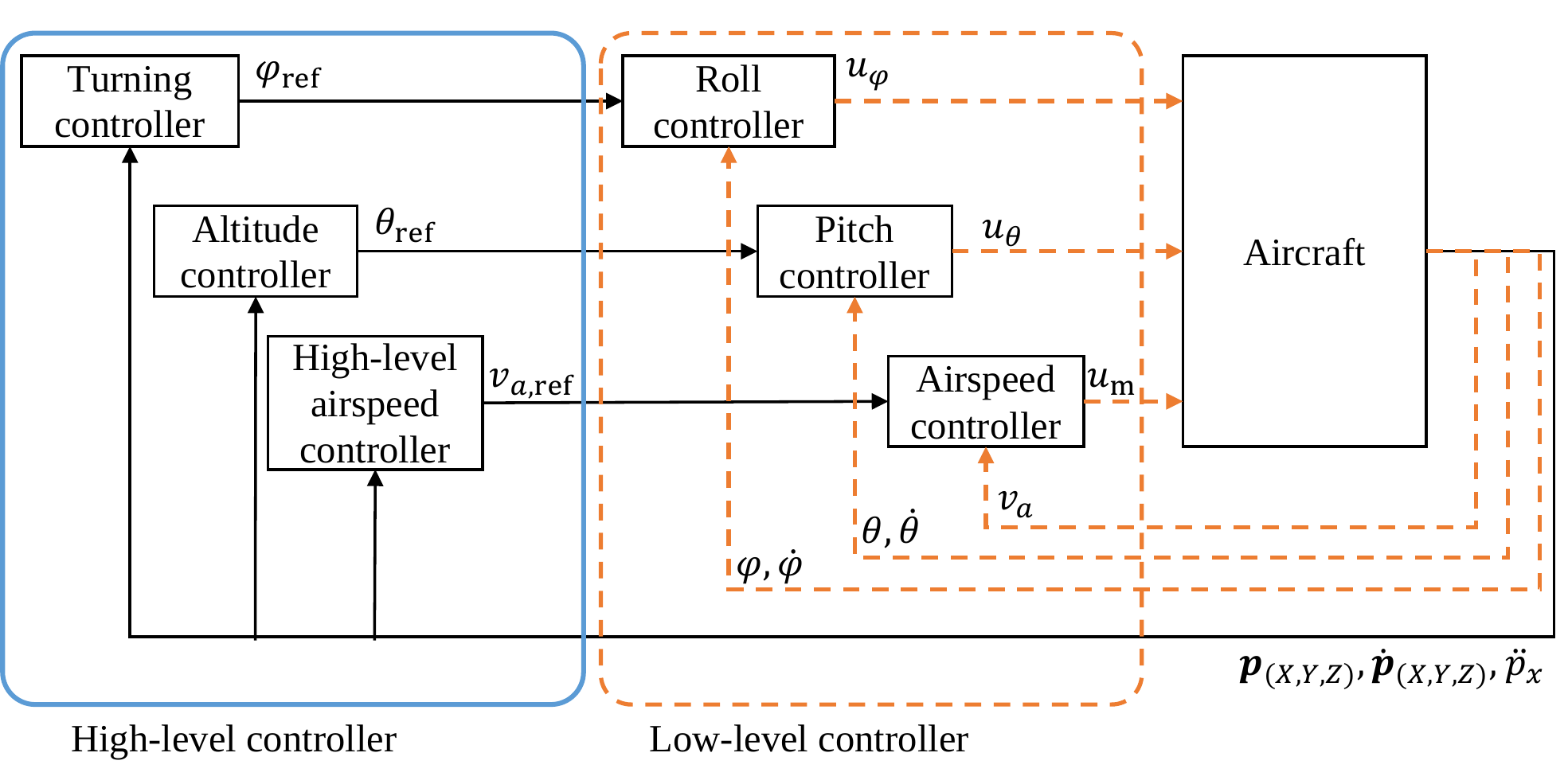}}
		\caption{Controller architecture showing the feedback variables used by the three low-level controllers and by the corresponding three high-level controllers.}\label{G:AutoControllerArchi}
	\end{figure}

	We employ a hierarchical control structure (see Fig~\ref{G:AutoControllerArchi}), with low-level controllers designed to track reference values for the roll angle, $\varphi_\text{ref}$, pitch angle, $\theta_\text{ref}$, and front airspeed, $v_{a,\text{ref}}$, and high-level controllers that compute such references in order to achieve the goals outlined in section \ref{SS:objectives} for each operational phase.
	Following our modeling approach, we design decoupled low- and high-level controllers for the roll, pitch and forward motion modes. The employed control inputs are those appearing in the model of section \ref{SS:model_id}, i.e. the aileron and elevator positions, $u_\varphi$ and $u_\theta$, and the propeller thrust, $u_\text{m}$. In the next sub-sections, we describe the structure of the low- and high-level controllers, followed by short discussions about the controllers' tuning and closed-loop stability guarantees.
	
	\subsubsection{Low-level controllers}\label{SSS:low_level}
	Since $\varphi,\,\dot{\varphi},\,\theta,\,\dot{\theta},$ and $v_a$ are directly measured and their dynamics are approximately linear for the sake of our goals (see section \ref{SS:model_id}), we can use static, linear state-feedback control laws to stabilize these quantities and track their reference values.
	
	For the roll dynamics, we define the angular tracking error as
	\begin{equation}\label{E:tracking_error_roll}
	e_\varphi(t)\doteq\varphi_\text{ref}(t)-\varphi(t),
	\end{equation}
	and employ the state-feedback control law
	\begin{equation}\label{E:fbk_roll}
	u_\varphi(t)=K_{e_\varphi}\,e_\varphi(t)+K_{\dot{e}_\varphi}
	\dot{e}_\varphi(t).
	\end{equation}
	From \eqref{E:roll} and \eqref{E:tracking_error_roll}-\eqref{E:fbk_roll} we have:
	\begin{equation}\label{E:roll_state_fbk}
	\begin{array}{r}
	\left[
	\begin{array}{c}
	\dot{e}_\varphi(t)\\\ddot{e}_\varphi(t)
	\end{array}\right]=\underbrace{\left[
		\begin{array}{cc}0&1\\-b_\varphi\,K_{e_\varphi}&a_\varphi-b_\varphi\,K_{\dot{e}_\varphi}
		\end{array}\right]}_{A_{\varphi,CL}}\left[
	\begin{array}{c}
	e_\varphi(t)\\\dot{e}_\varphi(t)
	\end{array}\right]\\
	+\left[\begin{array}{c}
	0\\w_\varphi(t)
	\end{array}\right]
	\end{array}
	\end{equation}
	where $w_\varphi(t)\doteq(\ddot{\varphi}_\text{ref}(t)-d_\varphi(t)-a_\varphi\dot{\varphi}_\text{ref})$ can be seen as an external, bounded disturbance term accounting for neglected dynamics and the time variation of the reference roll angle and rate issued by the high-level controller. We use a classical pole-placement technique \cite{Goodwin2001} to design the gains $K_{e_\varphi},\, K_{\dot{e}_\varphi}$ in order to assign the eigenvalues of the matrix $A_{\varphi,CL}$ describing the closed-loop dynamics of the tracking error. In particular, in order to easily tune the controller also during flight, we implemented the explicit equations that link $K_{e_\varphi},\, K_{\dot{e}_\varphi}$ to the desired closed-loop eigenvalues of the roll dynamics, denoted as $\lambda_{\varphi,1},\,\lambda_{\varphi,2}$:
	\begin{equation}\label{E:roll_gains_explicit}
	\begin{array}{rcl}
	K_{e_\varphi} &= & \dfrac{\lambda_{\varphi,1}\,\lambda_{\varphi,2}}{b_\varphi}\\
	K_{\dot{e}_\varphi} &= & \dfrac{\lambda_{\varphi,1}+\lambda_{\varphi,2}-a_\varphi}{-b_\varphi}\\
	\end{array}
	\end{equation}
	In \eqref{E:roll_gains_explicit}, the only design parameters are $\lambda_{\varphi,1},\,\lambda_{\varphi,2}$, which can be intuitively tuned on the basis of the desired closed-loop bandwidth (we provide the values employed in our experimental tests in section \ref{SS:implementation}).
	
	Following the same approach, we design the low-level controller for the pitch as:
	\begin{subequations}\label{E:pitch_control}
		\begin{gather}
		e_\theta(t)\doteq\theta_\text{ref}(t)-\theta(t)\\
		u_\theta(t)=K_{e_\theta}\,e_\theta(t)+K_{\dot{e}_\theta}
		\dot{e}_\theta(t)\label{E:fbk_pitch}\\
		K_{\dot{e}_\theta} =  -\dfrac{\lambda_{\theta,1}+\lambda_{\theta,2} -a_\theta}{-b_\theta}\\
		K_{e_\theta} = \dfrac{\lambda_{\theta,1}\,\lambda_{\theta,2}}{b_\theta}\\
		K_{\dot{e}_\theta} = \dfrac{\lambda_{\theta,1}+\lambda_{\theta,2}-a_\theta}{-b_\theta}
		\label{E:pitch_gains_explicit}
		\end{gather}
	\end{subequations}
	
	Finally, for the airspeed controller we employ a static law as well, however considering the square of the airspeed and of its reference:
	\begin{equation}\label{E:propeller_control}
	u_\text{m}(t)=K_\text{m}\left(v_{a,\text{ref}}^2(t)-v_a^2(t)\right),
	\end{equation}
	where $K_\text{m}>0$ is a design parameter.
	From \eqref{E:airspeed} and \eqref{E:propeller_control} we obtain the closed-loop relationship:
	\begin{equation}\label{E:propeller_closed_loop}
	v_a(t)=\sqrt{\dfrac{K_\text{m}}{K_\text{m}+\frac{1}{2}\rho A C_D}v_{a,\text{ref}}^2(t)+\dfrac{1}{K_\text{m}+\frac{1}{2}\rho A C_D}d_{v_a}(t)}
	\end{equation}
	which for large enough values of $K_\text{m}$ (as compared with the terms $\frac{1}{2}\rho A C_D$ and $d_{v_a}(t)$)  gives:
	\[
	v_a(t)\approx v_{a,\text{ref}}(t).
	\]
	
	As a final remark before proceeding to the description of the high-level controllers, we note that all the control inputs have hard saturations that need to be enforced, i.e. $u_\varphi\in[\underline{u}_\varphi,\,\overline{u}_\varphi]$, $u_\theta\in[\underline{u}_\theta,\,\overline{u}_\theta]$ and $u_\text{m}\in[\underline{u}_\text{m},\,\overline{u}_\text{m}]$. For the sake of simplicity, we did not include such saturations in the control laws \eqref{E:fbk_roll}, \eqref{E:fbk_pitch} and \eqref{E:propeller_control}, however in practice we trim the control inputs in order not to violate these bounds. Since we have no integrators in our controllers, the presence of the saturations does not give rise to windup issues. Moreover, as a matter of fact in our tests the saturation values were hit rarely and always for very short time, indicating that the available control authority was large enough to accomplish the targeted task. We provide the actual saturation values of our experimental setup in section \ref{SS:implementation}.
	
	\subsubsection{High-level controllers}\label{SSS:high_level}
	We adopt three high-level controllers for the steering, altitude and airspeed dynamics, which respectively compute the references $\varphi_\text{ref}(t)$, $\theta_\text{ref}(t)$, and $v_{a,\text{ref}}(t)$. While the low-level controllers never change during operation, we design different  high-level control strategies according to the operational phases outlined in section \ref{SS:objectives}, and we switch among them when a transition from the first phase to the second one is detected. Therefore, in the following we present the high-level controllers organized by operational phase, as well as the conditions that trigger the phase transition.
	
	\textbf{Take-Off and climbing}. In this phase, the aircraft is initially at rest on the slide, waiting for the acceleration provided by the latter. A minimum forward acceleration threshold $\underline{\ddot{p}}_{x,\text{to}}$ is set by the control designer. When the following condition is detected:
	\begin{equation}\label{E:acc_takeoff_condition}
	\ddot{p}_x\geq\underline{\ddot{p}}_{x,\text{to}}
	\end{equation}
	then a constant airspeed reference is issued:
	\begin{equation}\label{E:airspeed_takeoff}
	v_{a,\text{ref}}(t)=v_{a,\text{ref,to}},
	\end{equation}
	where the design parameter $v_{a,\text{ref,to}}$ is chosen as a much larger value than the cruise airspeed of the aircraft, in order to provide full propeller thrust during the initial climb. Regarding the altitude controller, a constant, relatively large reference pitch angle  $\theta_{\text{ref,to}}$ (e.g. 40$^\circ$) is used:
	\begin{equation}\label{E:pitch_takeoff}
	\theta_\text{ref}(t)=\theta_{\text{ref,to}}
	\end{equation}
	which gives place to a large vertical speed (see equation \eqref{E:vertical speed}). Finally, the roll reference is computed in order to keep a straight trajectory in the inertial $(X,Y)$ plane. To this end, we consider the course angle:
	\begin{equation}\label{E:course_angle}
	\gamma(t)=\arctan{\left(\dfrac{\dot{p}_Y(t)}{\dot{p}_X(t)}\right)}
	\end{equation}
	and we employ the following feedback controller:
	\begin{equation}\label{E:roll_takeoff}
	\varphi_\text{ref}(t)=K_{\varphi}\dfrac{|\dot{\boldsymbol{p}}(t)|}{g}\left(\gamma_{\text{ref,to}}-\gamma(t)\right),
	\end{equation}
	where $K_{\varphi}>0$ is a design parameter and $\gamma_{\text{ref,to}}$ corresponds to the orientation of the ground station's rails in the inertial reference frame. The latter can be easily and accurately estimated onboard by using the yaw angle provided by the IMU in a time interval before the initial acceleration, hence once again avoiding active communication between the ground station and the aircraft. The rationale behind equation \eqref{E:roll_takeoff} is the following: from \eqref{E:tracking_error_roll} we can write
	\begin{equation}\label{E:roll_cl_approx}
	\varphi(t)=\varphi_\text{ref}(t)-e_\varphi(t),
	\end{equation}
	where the tracking error $e_\varphi(t)$ can be considered to be small thanks to the low-level controller of the roll dynamics \eqref{E:fbk_roll}, particularly if we assume that the dominant poles of \eqref{E:roll_state_fbk} have been assigned such that the closed-loop roll motion is faster than the yaw motion (please see section \ref{SSS:tuning} for more details on the tuning and section \ref{SS:implementation} for the actual values used in our experiments). We further assume that
	\begin{equation}\label{E:yaw_rate_approx_course_angle_rate}
	\dot{\psi}(t)\simeq\dot{\gamma}(t),
	\end{equation}
	i.e. that the yaw rate and the rate of the course angle are similar, which is a reasonable assumption when the sideslip angle of the aircraft is small like in our application. Then, using \eqref{E:yaw_rate} and \eqref{E:roll_takeoff}-\eqref{E:yaw_rate_approx_course_angle_rate}, we have:
	\begin{equation}\label{E:yaw_cl_approx}
	\dot{\gamma}(t)\simeq K_{\varphi}\left(\gamma_{\text{to}}-\gamma(t)\right),
	\end{equation}
	which is (since $K_{\varphi}>0$)  a stable first-order system with time constant $\tau_\gamma=1/K_{\varphi}$. Thus, under the proportional control law \eqref{E:roll_takeoff}, the roll angle of the aircraft is controlled in order to track the desired course angle.
	Finally, we limit the turning radius of the aircraft to a minimum value $R_\text{min}$ by setting the following bounds on the reference yaw rate (see equation \eqref{E:curvature}):
	\begin{equation}\label{E:roll_constraints}
	-\dfrac{|\dot{\boldsymbol{p}}(t)|^2}{g\,R_\text{min}}\leq\varphi_\text{ref}(t)\leq\dfrac{|\dot{\boldsymbol{p}}(t)|^2}{g\,R_\text{min}}.
	\end{equation}
	Similarly to the other design parameters, in section \ref{SS:implementation} we provide the numerical value chosen for $R_\text{min}$.
	
	\textbf{Transition phase and figure-of-eight patterns}.
	When the ``safe-altitude'' condition:
	\begin{equation}\label{E:safe_Z}
	p_Z(t)\geq\underline{Z}
	\end{equation}
	has been reached, the transition phase begins. The goal is to steer the aircraft back towards the ground station and engage a repetitive figure-of-eight pattern roughly above it, while at the same time continuing to ascend to the target altitude $Z_\text{ref}>\underline{Z}$. To achieve this result, in the airspeed controller we set a constant reference airspeed $v_{a,\text{ref}}(t)=v_{a,\text{ref,flight}}$ equal to the cruise speed of the aircraft.
	
	About the turning controller, we employ the same control law \eqref{E:roll_takeoff}, but with a time-varying reference course angle $\gamma_{\text{ref}}(t)$ (which was set to the constant value $\gamma_{\text{ref,to}}$ during the climb) in order to steer the aircraft towards two switching target points, which are fixed w.r.t. the ground and suitably chosen to achieve the desired flight patterns. In particular, let us consider two target points $\boldsymbol{p}^\text{ I}$ and $\boldsymbol{p}^\text{ II}$ defined in the inertial $(X,Y,Z)$ plane. On the $(X,Y)$ plane, these points are computed in order to be symmetrical w.r.t. to the location of the ground station along the take-off direction  $\gamma_\text{to}$, and slightly shifted to one side along the direction perpendicular to $\gamma_\text{to}$. The altitude of the points is set equal to $Z_\text{ref}$. As an example and without loss of generality, if we assume that the ground station is located at the origin of the inertial frame and that $\gamma_\text{to}=0$ (i.e. the take-off is carried out along the $X-$direction), then a suitable choice of the target points is:
	\begin{equation}\label{E:target_points}
	\boldsymbol{p}^\text{ I}=\left[
	\begin{array}{c}
	\frac{\Delta X_{\text{ref}}}{2}\\
	Y_{\text{ref}}\\
	Z_\text{ref}\end{array}
	\right],\;\;
	\boldsymbol{p}^\text{ II}=\left[
	\begin{array}{c}
	\frac{-\Delta X_{\text{ref}}}{2}\\
	Y_{\text{ref}}\\
	Z_\text{ref}\end{array}
	\right],
	\end{equation}
	where $\Delta X_{\text{ref}}>0$ is the chosen distance between target points along the take-off direction. A graphical example of this type of parametrization of the target points (but for $\gamma_{\text{to}}\simeq 15^\circ$) is shown in the experimental results of Fig. \ref{F:AL&F_GPS_rel_posi}.
	
	At the start of the transition phase (i.e. when condition \eqref{E:safe_Z} is detected), the target point farthest away from the aircraft is chosen as the active one, denoted as $\boldsymbol{p}^\text{ a}$. Then, the reference course angle is computed as:
	\begin{equation}\label{E:ref_course_angle_flight}
	\gamma_{\text{ref}}(t)=\arctan{\left(\dfrac{p^\text{ a}_Y(t)-p_Y(t)}{p^\text{ a}_X(t)-p_X(t)}\right)},
	\end{equation}
	i.e. the course angle corresponding to a straight line connecting the current $(X,Y)$ position of the aircraft with that of the target.
	The switching of the active target point happens when the aircraft's $(X,Y)$ position surpasses the position of the current target point, after projecting both positions on a direction corresponding to the take-off course $\gamma_{\text{ref,to}}$. In the previous example \eqref{E:target_points} with $\gamma_{\text{ref,to}}=0$, the corresponding switching rule would be:
	\begin{equation}\label{E:target_point_switching}
	\boldsymbol{p}^\text{ a}(t)=\left\{
	\begin{array}{l}
	\boldsymbol{p}^\text{ I}\text{ if $p_X(t)<-\frac{\Delta X_{\text{ref}}}{2}+\delta_X$}\\
	\boldsymbol{p}^\text{ II}\text{ if $p_X(t)>\frac{\Delta X_{\text{ref}}}{2}-\delta_X$}\\
	\boldsymbol{p}^\text{ a}(t^-)\text{ else}
	\end{array}\right.
	\end{equation}
	where $\delta_X>0$ is a small tolerance (e.g. 0.5$\,$m) to avoid possible numerical issues when computing the $\arctan$ function in  \eqref{E:ref_course_angle_flight}, and $\boldsymbol{p}^\text{ a}(t^-)$ is the previous active target point (in the discrete-time implementation of the controller --see section \ref{SS:implementation}-- this corresponds to the target point at the previous sampling instant). The value of $\gamma_{\text{ref}}(t)$ computed by \eqref{E:ref_course_angle_flight} is finally plugged into \eqref{E:roll_takeoff} in place of $\gamma_{\text{ref,to}}$ to close the high-level turning control loop.
	
	About the altitude controller, in this phase the aim is to regulate the aircraft $Z$ position close to the reference $Z_\text{ref}$. To achieve this goal, differently from the initial take-off and climbing phase we employ a static proportional feedback controller to compute $\theta_\text{ref}(t)$:
	\begin{equation}\label{E:ref_theta_flight}
	\theta_\text{ref}(t) = \dfrac{K_{\theta}}{|\dot{\boldsymbol{p}}(t)|}\left(Z_{\text{ref}}-p_Z(t)\right)
	\end{equation}
	where $K_{\theta}>0$ is a design parameter. Similarly to what shown in \eqref{E:yaw_cl_approx} for the turning dynamics, from \eqref{E:pitch_control} we have:
	\[
	\theta(t)=\theta_\text{ref}(t)-e_\theta(t),
	\]
	with small tracking error $e_\theta(t)$ thanks to the low-level controller. Then, considering also \eqref{E:vertical speed} and \eqref{E:ref_theta_flight} we obtain:
	\begin{equation}\label{E:Z_cl_approx}
	\dot{p}_Z(t)\simeq K_{\theta}\left(Z_{\text{ref}}-p_Z(t)\right),
	\end{equation}
	which is a stable first-order system with time constant $\tau_\theta=1/K_{\theta}$.

	\subsubsection{Controller tuning and stability/robustness aspects}\label{SSS:tuning}
	
	The described controller features several tuning parameters, 
	which can be initially set on the basis of standard control design guidelines and physical insight, and then fine-tuned directly during experimental tests. We now provide for completeness some suggestions about possible tuning choices and procedures. The final numerical values employed in our tests are shown in section \ref{SS:implementation}
	
	The closed-loop eigenvalues  $\lambda_{\varphi,1},\,\lambda_{\varphi,2}$ of the low-level roll controller and the gain $K_{\varphi}$ of the high-level turning controller have to be chosen such that the bandwidth of the inner closed-loop dynamics \eqref{E:roll_state_fbk} is sufficiently larger than the one of the outer loop \eqref{E:yaw_cl_approx}, i.e.:
	\begin{equation}\label{E:roll_tuning}
	|\lambda_{\varphi,1}|,\,|\lambda_{\varphi,2}|> K_{\varphi},
	\end{equation}
	and similarly for the pitch/altitude controllers:
	\begin{equation}\label{E:pitch_tuning}
	|\lambda_{\theta,1}|,\,|\lambda_{\theta,2}|> K_{\theta}.
	\end{equation}
	The low-level controllers can be fine-tuned in real-world tests by using a fly-by-wire mode, where a pilot on the ground issues reference roll and pitch values to the aircraft and adjusts the values of the closed-loop eigenvalues. When the latter have been fixed, the gains of the outer loop can be also fine-tuned during fully autonomous flight tests.
	
	The airspeed controller gain $K_\text{m}$ can be initially set in order to provide full motor power (resp. stop the motor) when the measured airspeed is larger (resp. smaller) than the reference by a certain amount, e.g. $30\,\%$, and then fine-tuned during experiments.
	
	The acceleration threshold $\underline{\ddot{p}}_{x,\text{to}}$ should be chosen in order to be sure to detect the slide movement, while avoiding false-positives e.g. due to vibrations induced by wind gusts. The actual acceleration experienced by the glider depends mainly on the rails' length and on the take-off speed (assuming that all of the available length is exploited in order to limit the peak power consumed during take-off). As an example, in our tests the initial acceleration peak was about $40\,$N/m$^2$, and a threshold of $20\,$N/m$^2$ was large enough to avoid false-positives.
	
	As already mentioned, the values of $v_{a,\text{ref,to}}$ and $\theta_{\text{ref,to}}$ shall be chosen large enough to provide full motor thrust (taking into account the value of $K_\text{m}$) and a large climb angle during the take-off, in order to quickly gain altitude and increase the relative angle between the tether and the aircraft's velocity vector, hence reducing the risk of stall due to the tether's pulling force. The ``safe-altitude'' value $\underline{Z}$ shall be chosen to have enough safety margin in terms of distance from the ground, while avoiding to climb for too long on a straight line, with consequent large values of unreeled tether length.
	
	Finally, the minimum radius $R_\text{min}$, the cruise airspeed $v_{a,\text{ref,flight}}$, and the position of the target points $\boldsymbol{p}^\text{ I}$ and $\boldsymbol{p}^\text{ II}$ shall be chosen in order to keep the resulting flight paths compact (to reduce the occupied air volume and the length of the tether) but at the same time avoiding too large roll angles during the turns, taking into account the aerodynamic/manoeuvrability features of the employed aircraft.
	
	Before proceeding to the experimental results, we finally comment on the closed-loop stability  and robustness of the proposed control approach. Regarding nominal stability (i.e. assuming that the model equations introduced in section \ref{SS:model_id} hold with no uncertainty or disturbances), the choice of closed-loop eigenvalues with negative real-part for the low-level controllers and of positive gains $K_\text{m},\,K_{\varphi}$ and $K_\theta$ is sufficient to guarantee that the closed-loop system is asymptotically stable, which implies that bounded exogenous perturbations (e.g. due to the wind or the tether) give rise to bounded tracking errors. Moreover, since all of the feedback variables are directly measured, i.e. there is no observer involved, the use of state-feedback controllers provides good robustness margins against neglected dynamics, couplings between the different modes (which are neglected in our model) and parametric uncertainty \cite{Goodwin2001}. Thus, the control system can tolerate quite large inaccuracies in the estimate of all of the unknown parameters in the model (i.e. $a_\varphi,\,b_\varphi,\,a_\theta,\,b_\theta,\,C_D$, see section \ref{SS:model_id}) without compromising stability and performance.
	
	Albeit these few comments do not represent a rigorous robustness analysis, which lies beyond the scope of this paper, they provide some insight on why a relatively simple approach like the one presented here provides good results in real-world experiments, as shown in the next section.

	\section{Experimental results}\label{S:results}
	
	\subsection{Controller implementation, system and design parameters}\label{SS:implementation}
	
	We implemented the described controller on the Arduino MEGA 2560 board  installed on the glider. We employed a sampling frequency of $50\,$Hz, which is about two orders of magnitude larger than the fastest closed-loop dynamics of the system. Since the controller features only static feedback functions, there is no modification needed due to the discretization. In addition to the main control functions presented in the previous section, we programmed several consistency checks on the measured variables, in order to detect a possible sensor malfunctioning, as well as safety measures, like an automatic tether detach procedure if the distance of the aircraft from the ground station is too close to the maximum tether length (150 m). Since these additional functionalities do not change the system behavior in normal conditions, we omit further details here.
	
	The model parameters (resulting either from the identification procedure described in section \ref{SS:model_id} or from prior knowledge of the aircraft) are shown in  Table \ref{T:Numerical_value}, together with the employed values of the controller's parameters. We followed the tuning guidelines described in the previous section, fine-tuned the parameters during initial flight tests and then fixed them to the values of Table \ref{T:Numerical_value} for all the subsequent experiments.
	
	\begin{table}[!htb]
		\caption{System and controller parameters for the experimental tests}
		\label{T:Numerical_value}
		\centering
		\begin{tabular}{|l|r|l|}\hline
			\multicolumn{3}{|l|}{\textbf{System parameters}}\\\hline\hline
			$a_{\varphi}$ & $-2.3$ & s$^{-1}$\\\hline
			$b_{\varphi}$ & $12.6$ & s$^{-2}$\\\hline
			$a_{\theta}$ & $-4.65$ & s$^{-1}$\\\hline
			$b_{\theta}$ & $30$ & s$^{-2}$\\\hline
			$C_D$ & $0.05$ & -\\\hline
			$\rho$ & $1.2$ &kg$\,$m$^{-3}$\\\hline
			$A$ & $0.3$ & m$^{2}$\\\hline\hline
			\multicolumn{3}{|l|}{\textbf{Controller parameters - low level}}\\\hline\hline
			$\lambda_{\varphi,1}$ & $-2.7$ & s$^{-1}$\\\hline
			$\lambda_{\varphi,2}$ & $-3.1$ & s$^{-1}$\\\hline
			$\lambda_{\theta,1}$ & $-2.7$ & s$^{-1}$\\\hline
			$\lambda_{\theta,2}$ & $-3.1$ & s$^{-1}$\\\hline
			$K_{\text{m}}$ & $0.5$ & kg$\,$m$^{-1}$\\\hline
			$[\underline{u}_\varphi\;\overline{u}_\varphi]$& $[-0.34\;0.34]$&rad\\\hline
			$[\underline{u}_\theta\;\overline{u}_\theta]$ &$ [-0.34\;0.34]$& rad\\\hline
			$[\underline{u}_\text{m}\;\overline{u}_\text{m}]$&$ [0\;20]$& N\\\hline\hline
			\multicolumn{3}{|l|}{\textbf{Controller parameters - high level}}\\\hline\hline
			$K_\varphi$ & $1$ & s$^{-1}$\\\hline
			$K_\theta$ & $0.1$ & s$^{-1}$\\\hline
			$\underline{\ddot{p}}_{x,\text{to}}$& $20$ & m$\,$s$^{-1}$\\\hline
			$v_{a,\text{ref,to}}$& $16$ & m$\,$s$^{-1}$\\\hline
			$\theta_{\text{ref,to}}$& $0.69$ & rad\\\hline
			$\underline{Z}$& $20$ & m\\\hline
			$R_\text{min}$& $20$ & m\\\hline
			$v_{a,\text{ref,flight}}$& $13$ & m$\,$s$^{-1}$\\\hline
			$[p^\text{ I}_X\;p^\text{ I}_Y\;p^\text{ I}_Z]$&$ [30\;55\;50]$& m\\\hline
			$[p^\text{ II}_X\;p^\text{ II}_Y\;p^\text{ II}_Z]$&$ [-30\;40\;50]$& m\\\hline
		\end{tabular}
	\end{table}
	
	\subsection{Results}\label{SS:results}
	\begin{figure}[!hbt]
		\centerline{
			\begin{tabular}{c}
				(a)\\			\includegraphics[width=9cm]{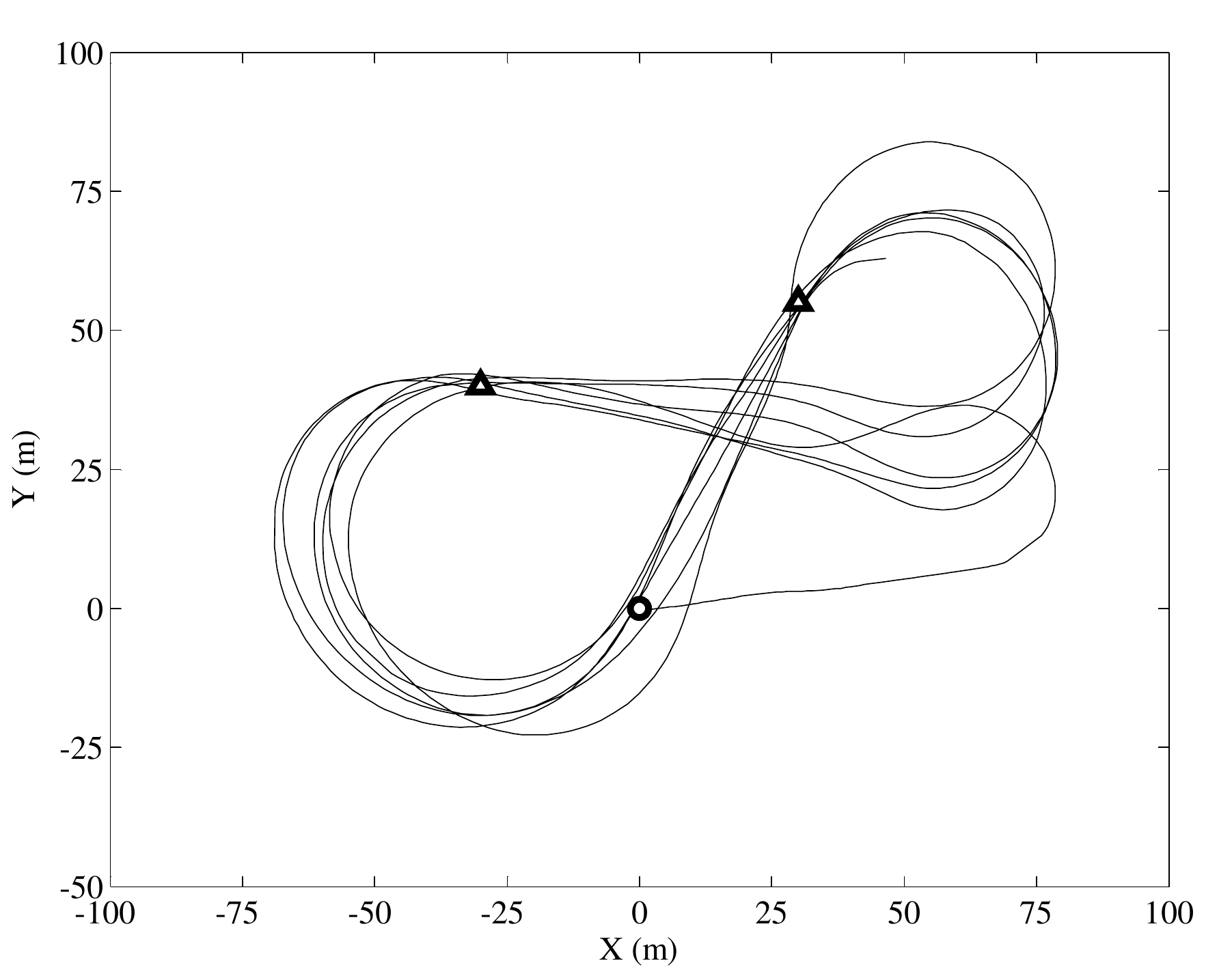}\\
				(b)\\
				\includegraphics[width=9cm]{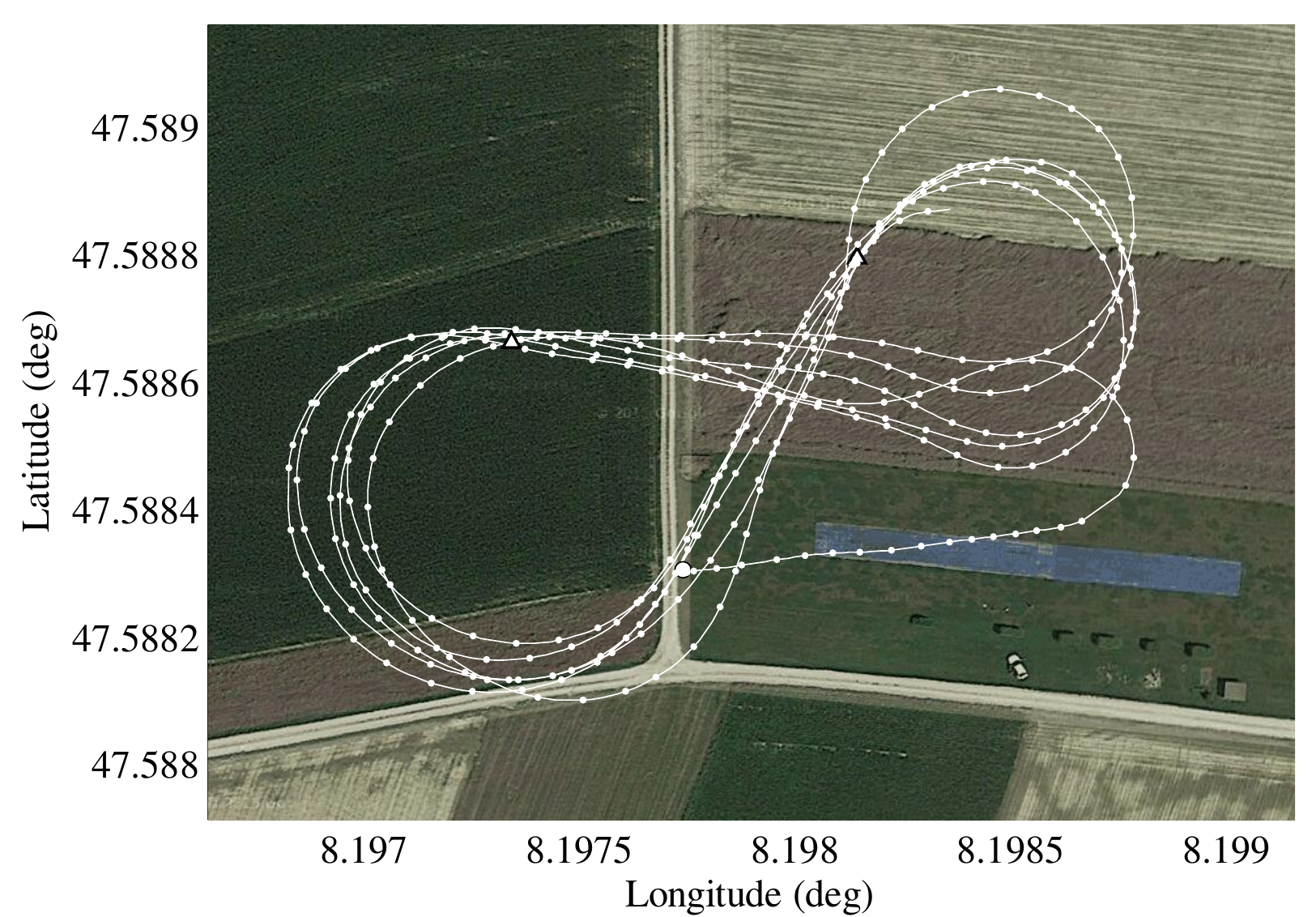}
			\end{tabular}
		} \caption{Experimental results of a test with little wind at ground level. (a) Typical flight pattern in the inertial $(X,Y)$ plane centered at the ground station (marked with '$\circ$'). The target points $\boldsymbol{p}^\text{ I}$ and $\boldsymbol{p}^\text{ II}$ are marked with '$\triangle$'. (b) Flight pattern in GPS coordinates, overlaid to a satellite map of the area. The white dots along the trajectory are position measurements down-sampled at $2\,$Hz.}\label{F:AL&F_GPS_rel_posi}
	\end{figure}
	\begin{figure}[!hbt]
		\centerline{
			\begin{tabular}{c}
				(a)\\			\includegraphics[width=9cm]{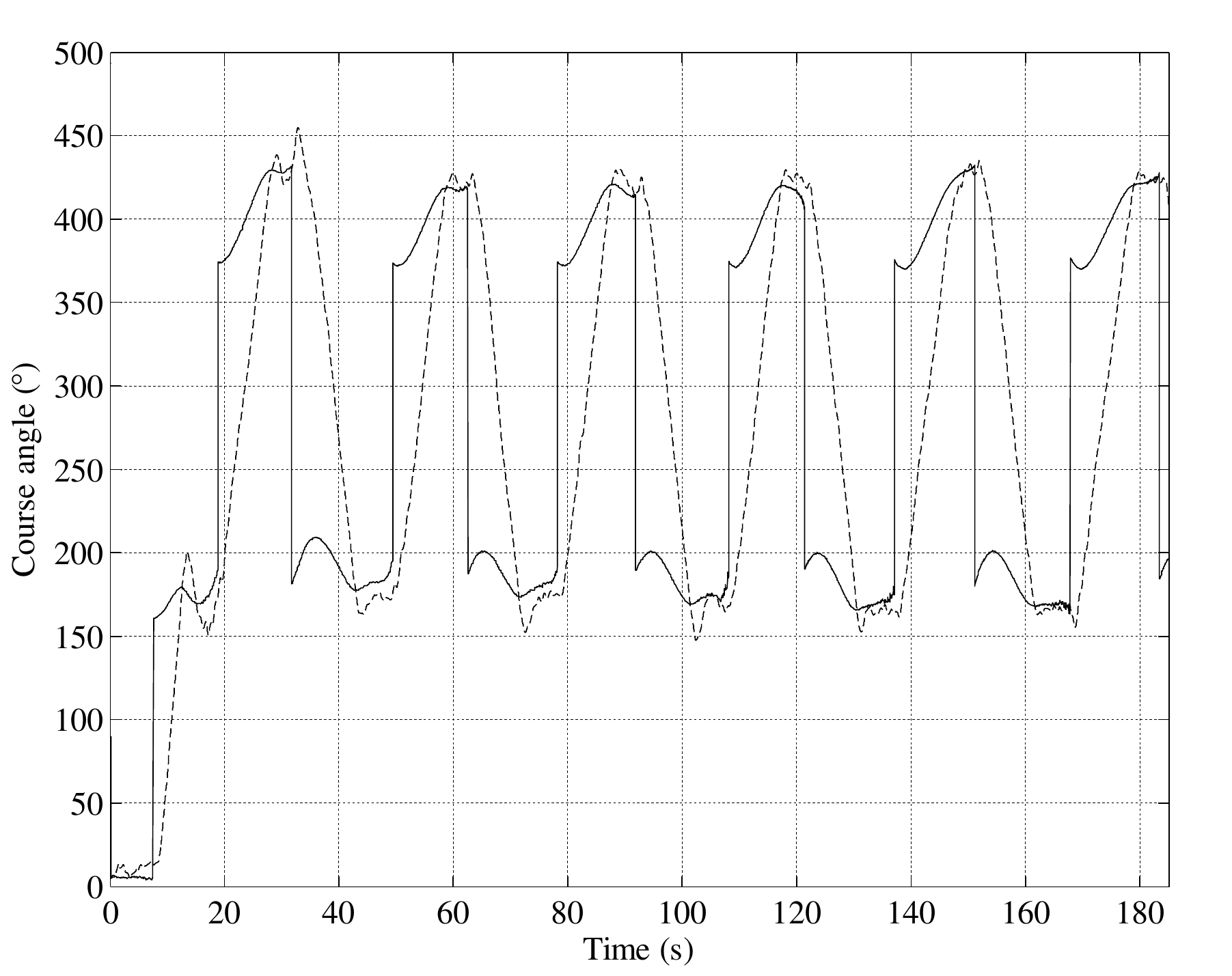}\\
				(b)\\
				\includegraphics[width=9cm]{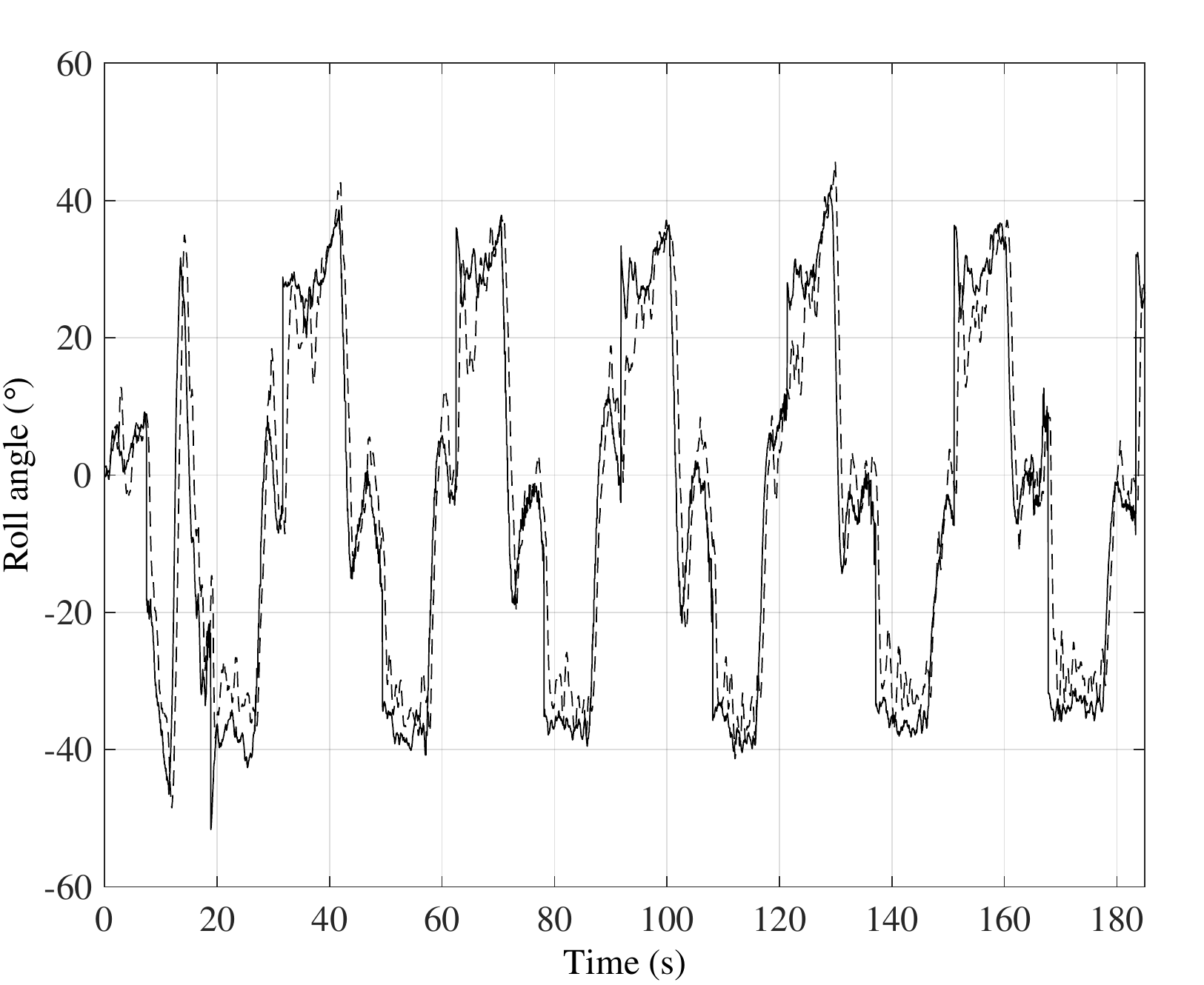}\\
				(c)\\
				\includegraphics[width=9cm]{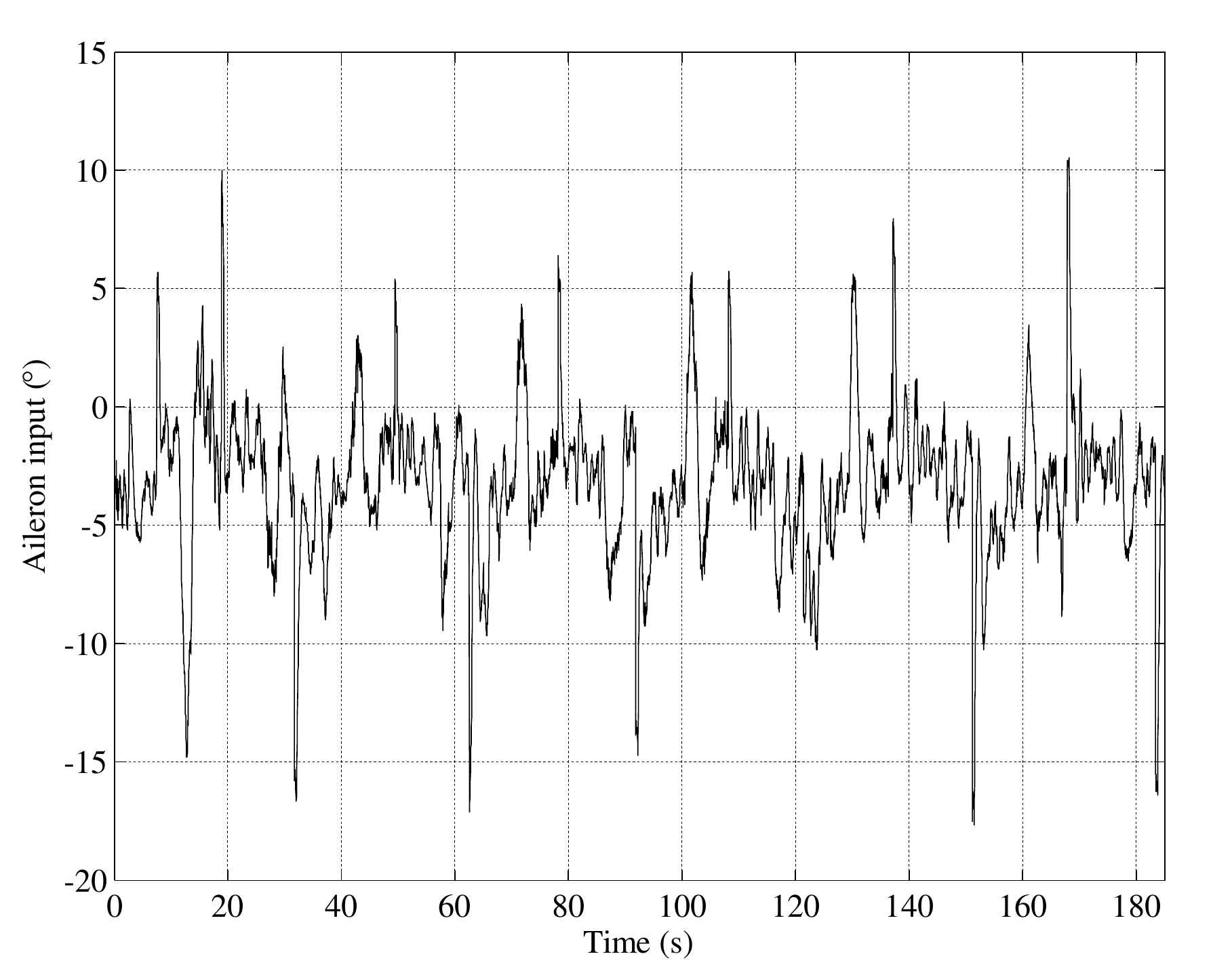}
			\end{tabular}
		} \caption{Experimental results corresponding to Fig. \ref{F:AL&F_GPS_rel_posi}, turning dynamics. (a) Course angle $\gamma$ (dashed) and its reference $\gamma_\text{ref}$ (solid) during tethered launch, climb and flight.  (b) Roll angle $\varphi$ (dashed) and its reference $\varphi_\text{ref}$ (solid) in the same test. (c) Aileron input $u_\varphi$.}\label{F:AL&F_Heading_Roll}
	\end{figure}
	
	\begin{figure}[!hbt]
		\centerline{
			\begin{tabular}{c}
				(a)\\			\includegraphics[width=9cm]{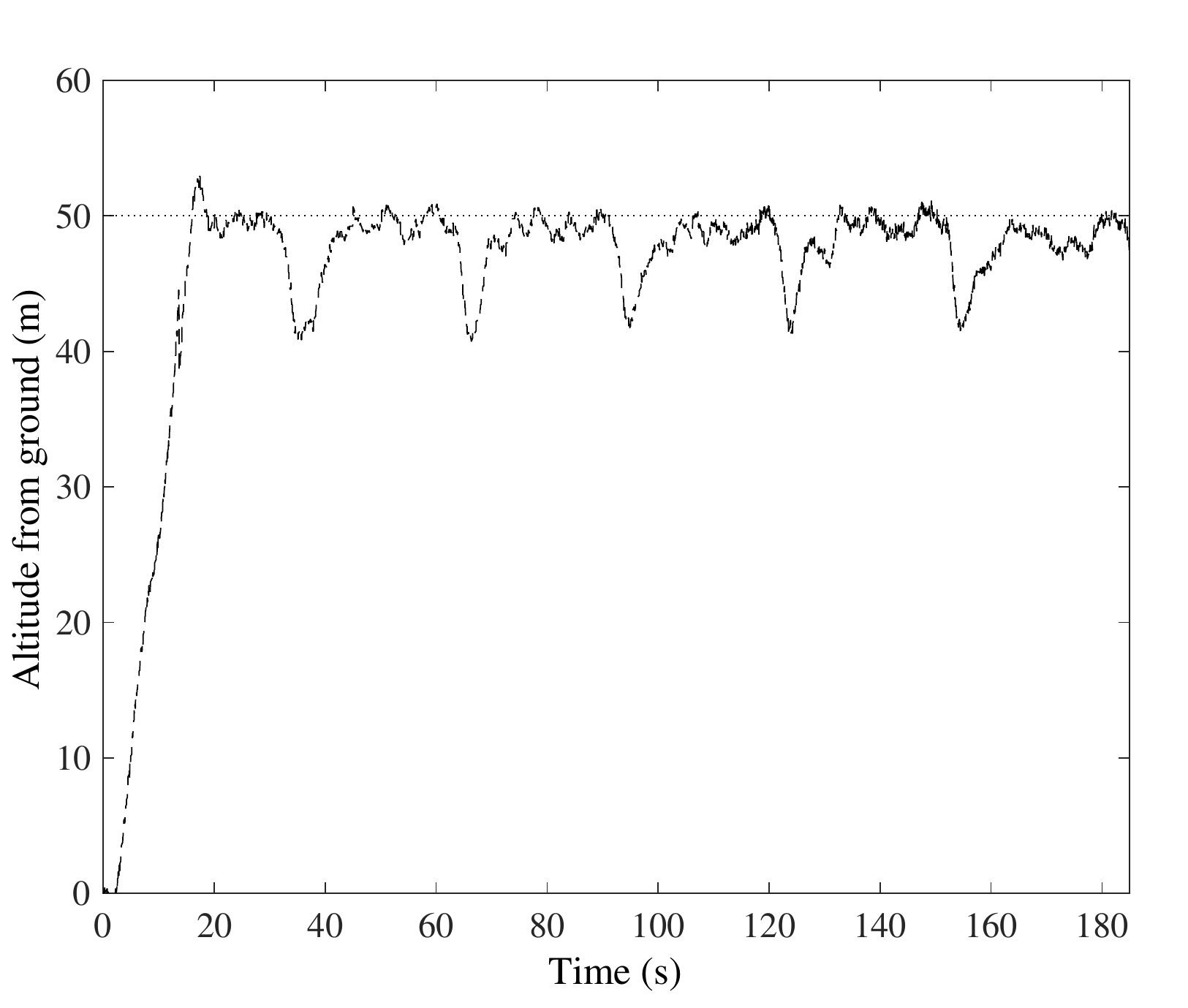}\\
				(b)\\
				\includegraphics[width=9cm]{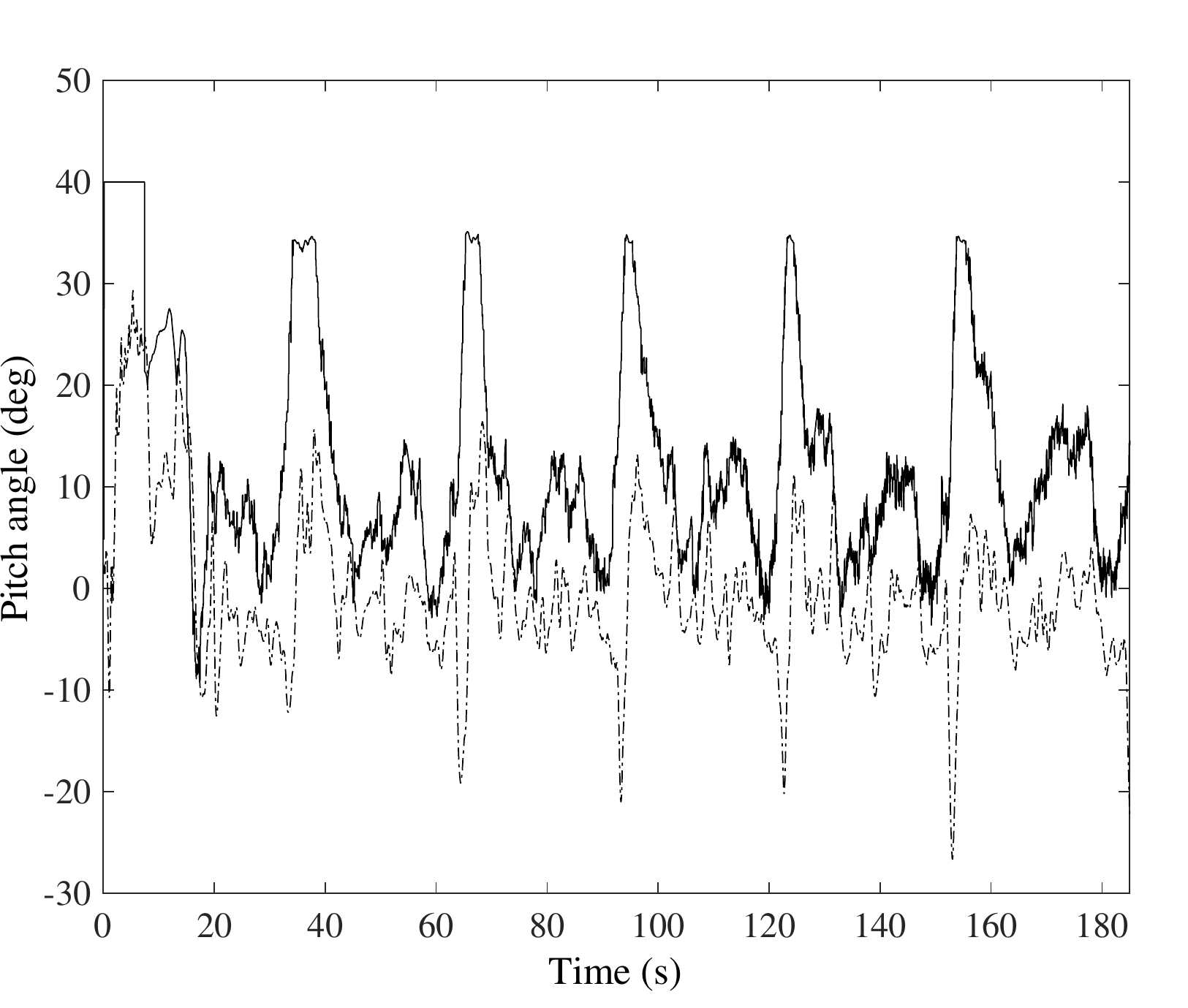}\\
				(c)\\
				\includegraphics[width=9cm]{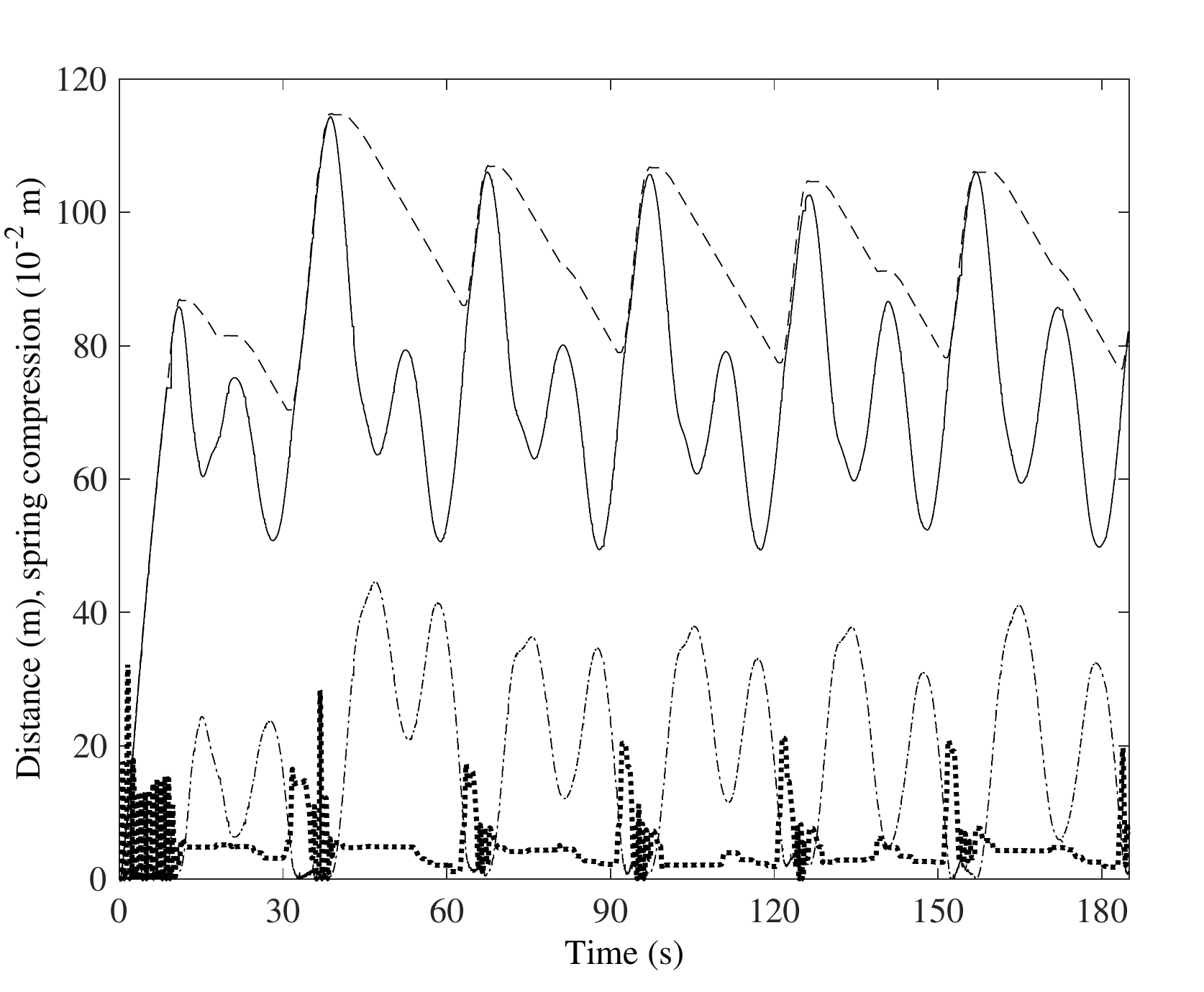}
			\end{tabular}
		} \caption{Experimental results corresponding to Fig. \ref{F:AL&F_GPS_rel_posi}, pitch and altitude dynamics. (a) Courses of the reference altitude $p_{Z,\text{ref}}$ (dotted line) and the actual one, $p_Z$ (dashed). (b) Courses of the reference pitch angle $\theta_\text{ref}$ (solid), and the actual one, $\theta$ (dash-dot). (c) Courses of the aircraft's distance from the ground station, $\|\boldsymbol{p}\|_2$ (solid line), of the tether length (dashed), of the spring compression (dotted) and of the slack tether length (dash-dotted). All plotted variables are in m, except for the spring compression in $10^{-2}\,$m.}\label{F:climb_path_to45m}
	\end{figure}
	
	We carried out several outdoor tests 
	using our experimental setup and the described control system. All the tests were carried out close to Leibstadt, Switzerland, at a location with little/moderate wind on average, with prevalent direction from East/North-East. We present here the experimental data of 14 tests carried out with the same take-off course angle $\gamma_{\text{to}}$. During the tests, the wind conditions changed from little/zero wind to front wind of about 4-5 m/s, to side wind of about 3-4 m/s. Thus, we are able to present results with both undisturbed conditions (i.e. either no wind or little front wind) and in presence of moderate wind disturbance coming from different directions. All these considerations hold for the wind at ground level, while at the target altitude (50$\,$m) a moderate wind and wind gusts were always present, of the order of 3-4 m/s as estimated by experienced model glider pilots. Since we had no equipment to measure the wind at altitudes higher than a few meters, we cannot provide more precise estimates of the wind conditions experienced by the glider after the very first take-off instants. A movie is available online \cite{Fagiano2016a}, showing an autonomous take-off and flight test with the initial take-off, the transition phase and finally the figure-of-eight flights.
	
	We first present the results of a test with little wind at ground level. 
	The flight trajectory projected on the inertial $(X,Y)$ plane centered at the ground station is shown in Fig. \ref{F:AL&F_GPS_rel_posi}(a), and the same trajectory overlayed on a satellite image of the corresponding GPS coordinates is presented in Fig. \ref{F:AL&F_GPS_rel_posi}(b). The initial take-off phase on a straight line is clearly visible, as well as the transition to the figure-of-eight paths.
	
	Figs.~\ref{F:AL&F_Heading_Roll}(a)-(c) show the behavior of the turning dynamics (roll and course angle controllers) during the test.
	The discontinuities in $\gamma_\text{ref}(t)$ (Fig.~\ref{F:AL&F_Heading_Roll}(a), solid line) correspond to the switching instants between the target points. It can be clearly seen that the actual course angle $\gamma(t)$ (Fig.~\ref{F:AL&F_Heading_Roll}(a), dashed line) changes at a constant rate during these transitions, which corresponds to the turning rate at cruise speed and with the minimum turning radius of 20$\,$m set in the controller. The corresponding roll angle $\varphi(t)$ (Fig.~\ref{F:AL&F_Heading_Roll}(b)) is approximately constant and equal to about $35^\circ$ during these turns. Finally, the employed roll input values $u_\varphi$ (Fig.~\ref{F:AL&F_Heading_Roll}(c)) are usually smaller than $10^\circ$ with few peaks at $15^\circ$, quite far from the saturation values of $\pm20^\circ$. Overall, the inner and outer loops for the roll and turning dynamics show very good tracking performance and provided a very consistent and robust behavior.
	
	The closed-loop altitude and pitch motions are shown in Fig~\ref{F:climb_path_to45m}(a)-(b). The aircraft altitude quickly approaches the target of 50 m during the initial take-off and transition. The vertical rate is constant because, similarly to what happens for the rate of the course angle, the aircraft sets to constant pitch and airspeed, which according to our model \eqref{E:vertical speed} yields a constant vertical position rate. During flight, the high-level controller is able to track with good accuracy (about 3-4 m of error) the target altitude, see Fig. \ref{F:climb_path_to45m}(a) e.g. between $45\,$s and $60\,$s, except for periodic, quick drops in the altitude which are limited to less than 10$\,$m of tracking error. A comparison between the distance of the aircraft from the ground station and the tether length, shown in Fig. \ref{F:climb_path_to45m}(c), reveals that such disturbances are due to the tether pull: during each figure-of-eight pattern the aircraft pulls on the tether for a fraction of the time and some length of slack line (also shown in Fig. \ref{F:climb_path_to45m}(c)) is created. After the initial transient, the behavior of such a slack line becomes periodic as well, since the overall system (aircraft and ground station) converges to a stable periodic motion. The length of slack line depends on how fast the winch reels-in the line compared to how fast the glider changes its directions and flies back towards the ground station during the figure-of-eight patterns. The compression of the spring installed on the ground station is depicted in \ref{F:climb_path_to45m}(c): we recall that this signal is used by the winch controller to regulate the winch reeling speed and decide whether to reel-out or reel-in (as briefly described in section \ref{SS:GS_control} and thoroughly presented in \cite{Fagiano2016submitted}).
	
	Overall, the presented results show that the altitude controller behaves as expected and as discussed in section \ref{SSS:tuning}: in nominal/unperturbed conditions the tracking performance is very good, and in the presence of external disturbances the system reacts with bounded tracking errors. Moreover, it is shown that there is quite some room for improvement, either by better tuning or more advanced control strategies of the ground station, or by allowing the aircraft and ground station to communicate 
	and optimize their behavior in a coordinated way.
	
\begin{figure}[!htb]
		\centerline{
			\includegraphics[width=9cm,clip]{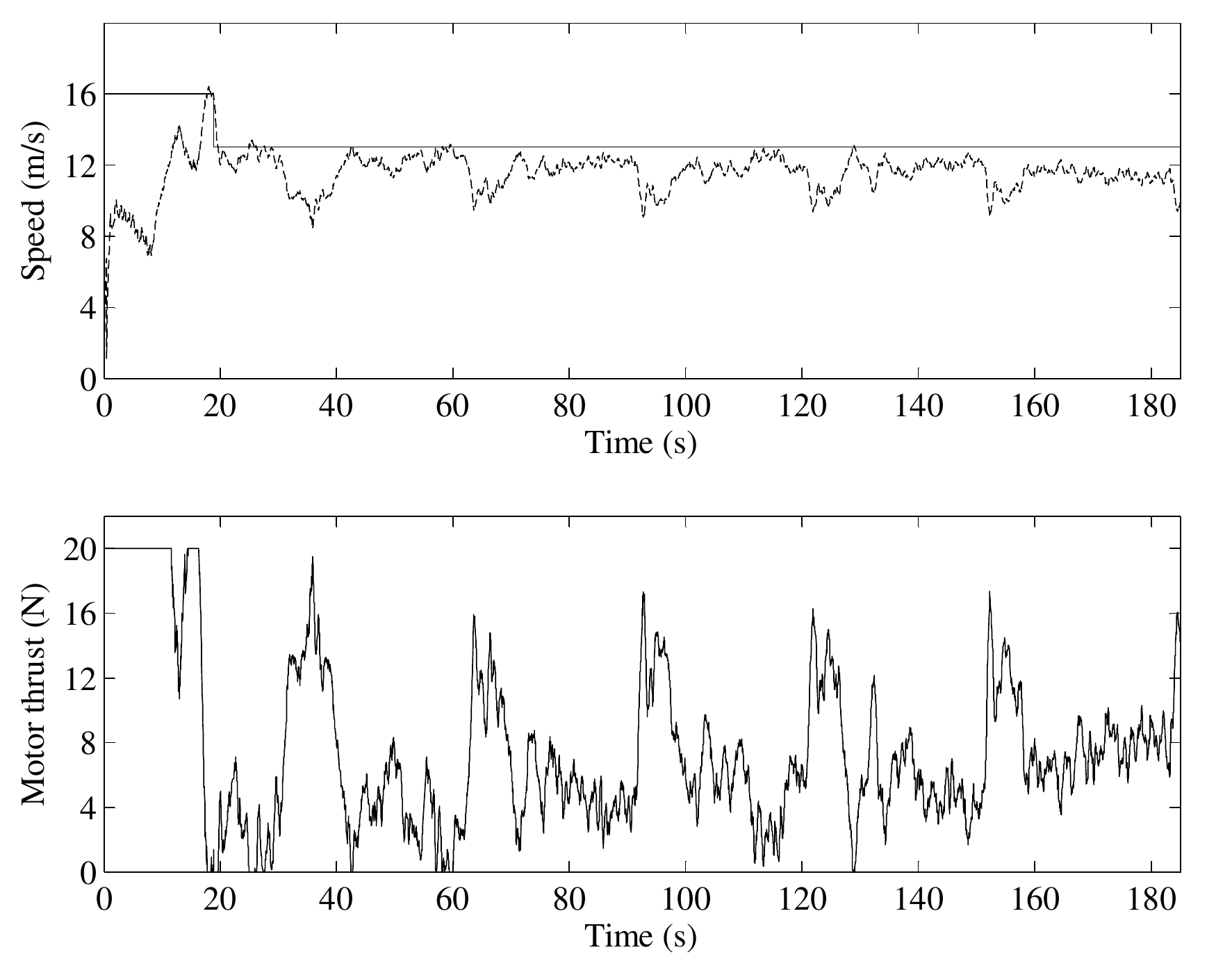}}
		\caption{Experimental results corresponding to Fig. \ref{F:AL&F_GPS_rel_posi}, airspeed and motor thrust. Upper plot: courses of the reference airspeed $v_{a,\text{ref}}$ (solid) and of the actual one, $v_a(t)$ (dashed). Lower plot: propeller thrust $u_\text{m}$.}\label{F:motorcontrol}
	\end{figure}

	
	Regarding the propeller controller, Fig. \ref{F:motorcontrol} shows the courses of the reference airspeed and of the actual one, as well as the corresponding propeller thrust $u_\text{m}$. Similarly to the pitch controller, with no tether perturbations a small tracking error of about 0.5 m$\,$s$^{-1}$ is achieved, while when the tether pulls on the aircraft a drop of airspeed is noticed, to which the controller reacts with an increase of thrust up to the maximum value. The loss of airspeed has consequences on all the other motions (pitch and roll), since it changes suddenly the angle of attack and the incoming flow speed. However, as already commented, the proposed controller is able to effectively cope with all such nonlinear effects.
	
\begin{figure}[h]
		\centerline{
			\includegraphics[width=9cm,clip]{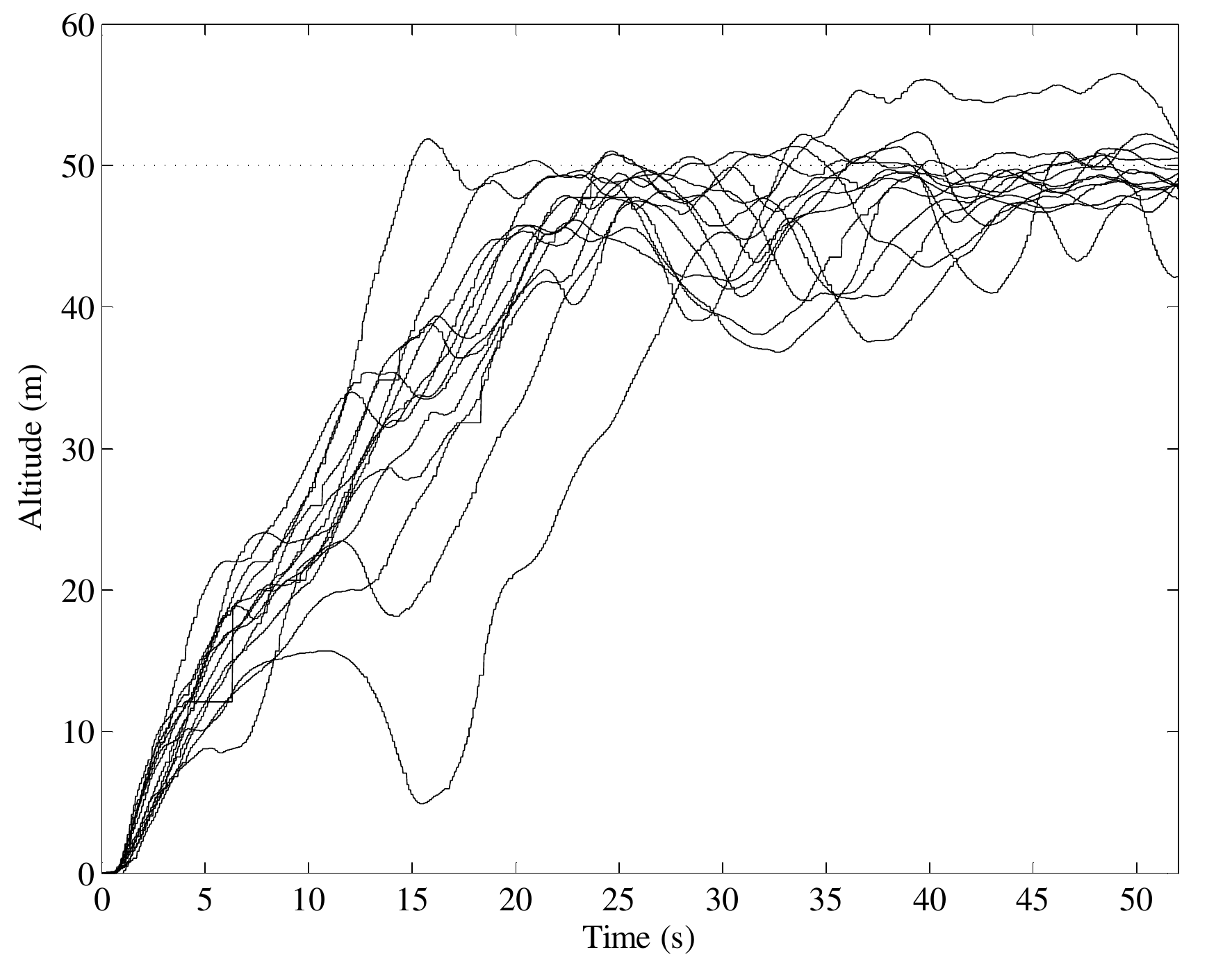}}
		\caption{Experimental results of 14 overlayed tests. Course of the aircraft altitude.}\label{F:climb_path_to45m_all}
	\end{figure}

\begin{figure}[h]
		\centerline{
			\includegraphics[width=9cm,clip]{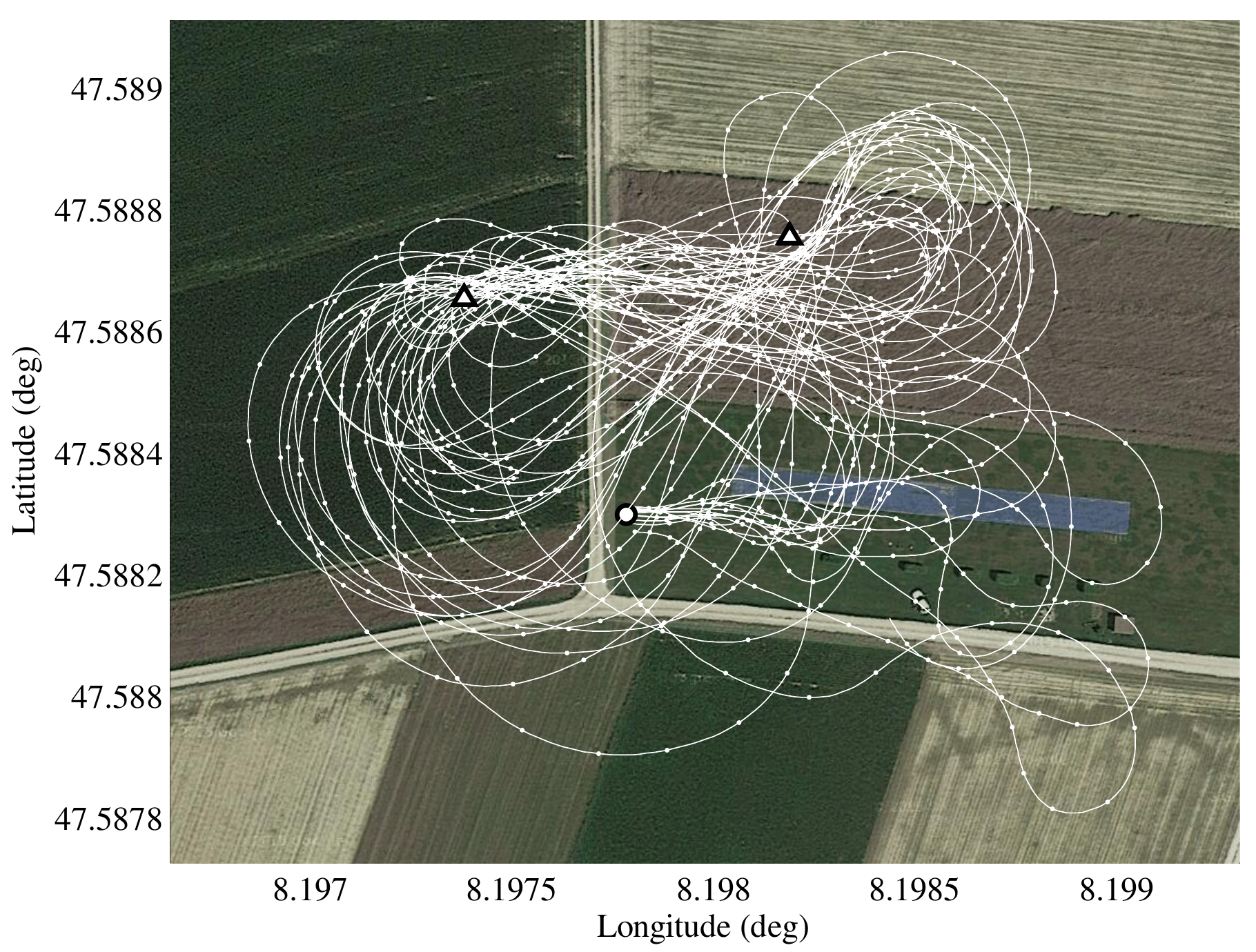}} 	\caption{Experimental results of 14 overlayed tests. Aircraft position in GPS coordinates. The white dots correspond to position measurements down-sampled to 1$\,$Hz. The ground station position is marked with '$\circ$'. The target points $\boldsymbol{p}^\text{ I}$ and $\boldsymbol{p}^\text{ II}$ are marked with '$\triangle$'.}\label{F:XY_path_all}
	\end{figure}

\begin{figure}[h]
		\centerline{
			\includegraphics[width=9cm,clip]{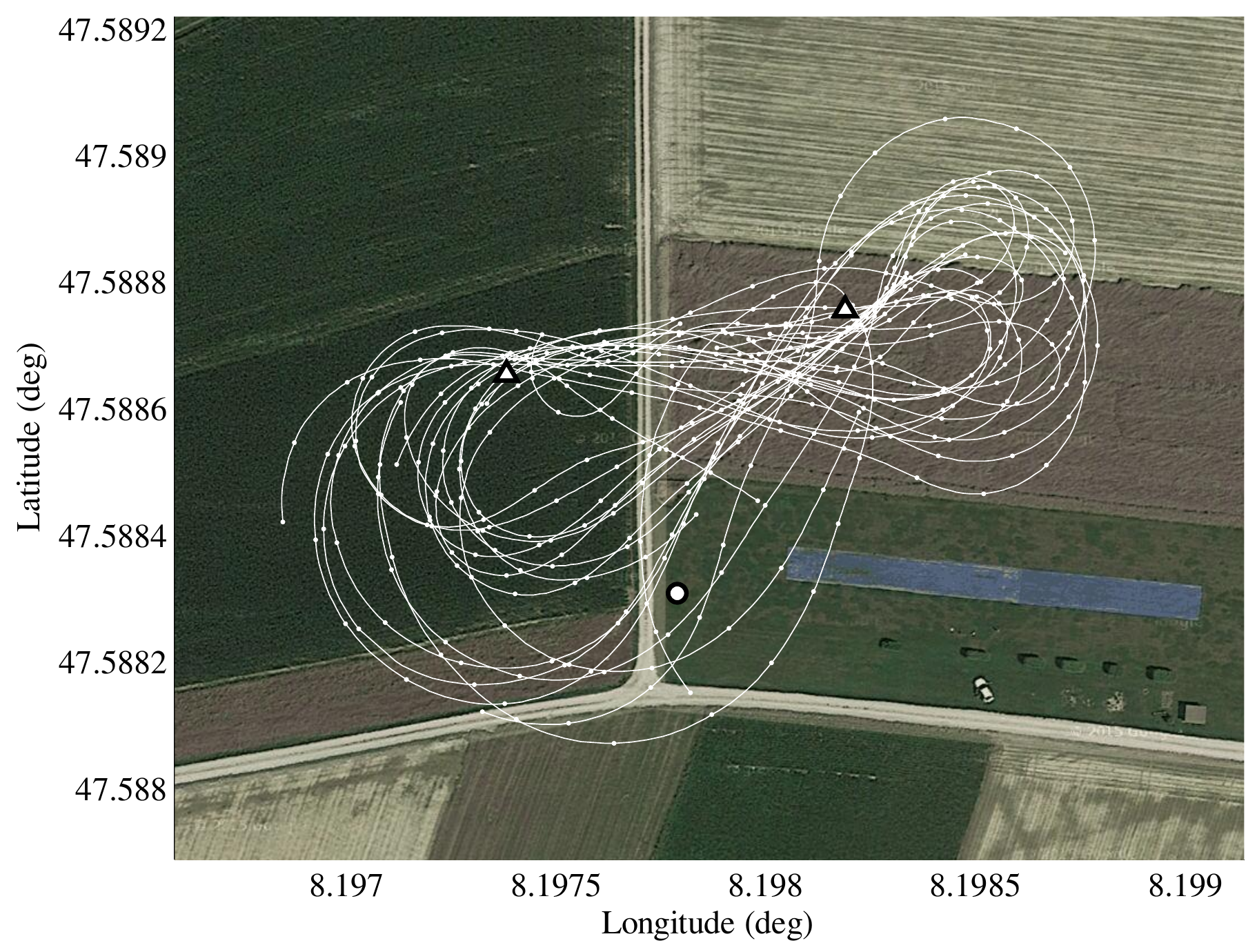}} 	\caption{Experimental results of 14 overlayed tests. Aircraft position in GPS coordinates without the initial take-off and transition phases. The white dots correspond to position measurements down-sampled to 1$\,$Hz. The ground station position is marked with '$\circ$'. The target points $\boldsymbol{p}^\text{ I}$ and $\boldsymbol{p}^\text{ II}$ are marked with '$\triangle$'.}\label{F:XY_path_all_no_transient}
	\end{figure}

	Finally, we present the results pertaining to all the 14 tests carried out with the same starting course angle, to show the repeatability and robustness of the controller in conditions with little to moderate wind disturbance. Fig. \ref{F:climb_path_to45m_all} shows the course of aircraft altitude for all the tests: it can be seen that the trajectories are quite consistent, except for one occurrence. The aircraft position on the $(X,Y)$ plane (in GPS coordinates) for all the tests is shown in Fig. \ref{F:XY_path_all}: there is  some variability in the initial transient and in the first few figure-of-eight patterns, until the aircraft and the whole system settle to a periodic motion. After such a transient, the patterns are more consistent but still exhibit some variability, which is expected: since the high-level controller relies on a two-point strategy only (instead for example on the tracking of a fully prescribed orbit), the actual obtained patterns can change e.g. depending on the wind conditions, similarly to what happens for the two-point strategies used for crosswind kite control in e.g. \cite{Fagiano2014} and \cite{Erhard2015}. However, as shown in Fig.  \ref{F:XY_path_all_no_transient} where the $(X,Y)$ trajectories are presented without the initial transient, in all the performed tests the control system was able to reach and consistently follow the figure-of-eight paths, which was the ultimate intended goal of our feasibility study.
	
	\section{Conclusions and future work}\label{S:conclusions}
	
	We presented a feedback controller for a tethered aircraft, able to consistently carry-out tethered take-off and low-force flight maneuvers. The controller works in combination (yet without any exchange of information) with a ground station that manages the initial acceleration on ground and controls the tether reeling, in order to limit the pulling force. The presented approach includes both modeling and control design aspects and has been extensively tested in experiments, whose results have been reported here as well.
	
	Our experimental results lead to essentially two concluding remarks. The first one is that a rather simple control approach, based on proportional controllers only and neglecting system nonlinearities, couplings among the aircraft's modes and the presence of the tether, is able to achieve satisfactory results in terms of performance and robustness. This indicates that the considered take-off approach appears to be technically viable and, on the basis of our previous studies \cite{Fagianoa}, it has the potential to also be economically viable, since it requires relatively small additional power both on ground and onboard and is able to carry out the take-off in very compact space. In our small-scale prototype, less than 2$\,$m of rails' length were enough to take-off. According to \cite{Fagianoa}, with a larger wing loading the required take-off distance would still remain of the order of the aircraft's wingspan, similarly to other AWE technologies which employ e.g. vertical take-off approaches, like Makani Power \cite{VanderLind2013}.
	
	The second consideration is that there is quite some room for improvement of the controller, along the following (and possibly more) lines of further research: advanced multi-variable controllers taking into account the couplings among the aircraft's modes; nonlinear control approaches; adaptive approaches based on the estimation of the wind conditions and on the aircraft's angle of attack and sideslip; coordinated control of the aircraft and of the ground station; iterative learning approaches; improved/specialized aircraft design also considering the presence of the tether.
	
	As a final remark, we believe that the next major technical bottleneck for the considered AWE technology, on which future research efforts should be targeted, is the feasibility proof (theoretical and experimental) of an autonomous landing strategy with the same qualitative  features as the take-off approach that we demonstrated in this paper, i.e. small required land area, low cost, repeatability and robustness. It may well be that the removal of such a bottleneck entails not only control technology, but also the development of new aircraft concepts specifically designed for this purpose.
	
	\section*{Acknowledgment}
	The authors would like to thank Alessandro Lauriola and Stefan Schmidt for their helpful contributions during the project.
	
	\bibliographystyle{plain}

\end{document}